\documentclass[prb,twocolumn,english,superscriptaddress, floatfix,longbibliography]{revtex4-2}
\usepackage{amsmath}
\usepackage{graphicx}
\usepackage{CJK}
\usepackage{amsgen}
\usepackage{amsfonts}
\usepackage{amsbsy}
\usepackage{amssymb}
\usepackage{dcolumn}
\usepackage{bm}
\usepackage{epsfig}
\usepackage{color}
\usepackage{lineno}
\usepackage{booktabs}
\usepackage{orcidlink}
\usepackage{xcolor}

\input{tcilatex}
\begin{document}

\title{Navigating entanglement via Ruderman-Kittel-Kasuya-Yosida exchange:
Oscillatory, boundary-residing, pulsed, and damping-stabilized trajectories}




\begin{CJK*}{UTF8}{bsmi}

\author{Son-Hsien Chen~\thanks{*Corresponding Author}~\orcidlink{0000-0002-3700-0018}}
\email{sonhsien@utaipei.edu.tw}
\affiliation{Department of Applied Physics and Chemistry, University of Taipei, Taipei 100234, Taiwan}

\author{Seng Ghee Tan~\orcidlink{0000-0002-3233-9207}}
\affiliation{Department of Optoelectric Physics, Chinese Culture University, Taipei 11114, Taiwan}

\author{Ching-Ray Chang~\orcidlink{0000-0003-1974-9583}}
\affiliation{Quantum Information Center, Chung Yuan Christian University, Taoyuan, 320314, Taiwan}




\date{\today}

\begin{abstract}
Entanglement dynamics are fundamental to quantum technologies, yet controlling their temporal evolution in a reversible and stable manner remains challenging. We propose a solid-state framework based on the Ruderman-Kittel-Kasuya-Yosida interaction, realizable in gate-defined quantum dots or suspended structures, in which two spin qubits couple to a central spin qudit that mediates an effective, time-dependent exchange. The dynamics are governed by an exchange-time integral that unifies interaction strength and physical time into a single scalar control variable, enabling \emph{time-reversible} and cyclic navigation of the Hilbert space. Crucially, we show that out-of-phase modulation grants access to higher entanglement subspaces, while introducing damping to the exchange modulation achieves stabilized trajectories that drive the system toward stationary entanglement values. This framework provides a systematic route for shaping entanglement dynamics, particularly in the \emph{near-boundary regime}, using exchange control alone, overcoming the limitations of monotonic evolution and offering practical strategies for entanglement stabilization in realistic solid-state architectures, with direct relevance to quantum metrology and environment-assisted entanglement engineering.
\end{abstract}


\maketitle

\end{CJK*}

\section{Introduction}

\label{sec:intro} 
Quantum entanglement refers to the nonclassical correlations between
subsystems, where the overall quantum state cannot be factored into
independent states of the individual parts~\cite%
{ifmmodeZelseZfiyczkowski2001,Horodecki2009}. This correlation plays a
pivotal role in emerging quantum technologies~\cite{Bennett2000},
underpinning advances in gravitational wave detection~\cite%
{Khalili2018,Zeuthen2019}, quantum cryptography~\cite%
{Ekert1991,Shor2000,Gisin2002,Bennett1998}, and quantum computation~\cite%
{Horodecki2009,DiVincenzo1995,Steane1998,DiVincenzo2000,Ladd2010,Nielsen2010,Madsen2022}%
. Despite its foundational importance, entanglement generation and the
control of its dynamics remain central challenges in quantum information
science~\cite{Preskill2018,RieraSabat2023,Zhang2024,Wang2025}.

To address these challenges, diverse quantum processor platforms have been
developed, including superconducting circuits~\cite{Arute2019}, trapped ions 
\cite{Cirac1995,Monroe2014}, and photonic qubits~\cite{Zhong2020}. Among
these, spin qubits in solid-state systems are particularly promising
candidates, specifically those based on magnetic impurities or defect
centers (e.g., nitrogen-vacancy centers in diamond, donor spins in silicon)
and lithographically or gate-defined quantum dots (QDs). Defect-based qubits
combine optical initialization and readout with long coherence times,
enabling remote entanglement over distances of up to two meters~\cite%
{Humphreys2018}. Donor spins in silicon achieve coherence times on the order
of seconds with gate fidelities above 99\%~\cite%
{Tyryshkin2012,Pla2013,Watson2018}, whereas QD spin qubits support dense
integration and fast all-electrical control, with resonant CNOT gate
fidelities above 98\%~\cite{Loss1998,Nowack2007,Petta2005,Zajac2018}. Both
architectures allow electrically tunable exchange coupling for rapid
two-qubit gates and coherence protection via dynamical decoupling~\cite%
{Petta2005,Bluhm2011}. This gate-voltage tunability enables adjustment of
the final entanglement over a broad range of strengths~\cite%
{Chen2024,Lin2025}.

However, realizing a scalable architecture with these spin qubits requires a
robust long-range coupling mechanism, as direct exchange interaction is
limited to nearest neighbors. To this end, both QDs~\cite%
{Vavilov2005,Petersson2012,Yang2016,Wang2022,Vonhoff2022,Utsumi2004,Stocker2024}
and magnetic impurities~\cite%
{Chen2009a,Allerdt2015,Mousavi2021,Kettemann2024} can couple via
Ruderman-Kittel-Kasuya-Yosida (RKKY) exchange~\cite%
{Ruderman1954,Kasuya1956,Yosida1957}. This interaction is mediated by the
host electrons, alternating between ferromagnetic and antiferromagnetic
regimes with distance, which facilitates entanglement generation~\cite%
{Cho2006,Allerdt2015}, even at long range~\cite{Yang2016,Elman2017}.

Environment-mediated entanglement, such as that induced by RKKY, exhibits
distinctive temporal behavior. It can change abruptly, leading to
entanglement sudden death (ESD)~\cite{Yu2004a,Yu2006a,Yu2006b,Ann2007,Yu2009}%
---a complete loss of entanglement in finite time---and its counterpart,
entanglement sudden birth (ESB)~\cite{Yuan2007,Wang2018,Chen2024}. These
phenomena, observed in various solid-state~\cite{Wang2018,Chen2024,Chen2025}
and optical systems~\cite{Almeida2007}, involve ESD-ESB transitions that can
be of finite (TFD) or zero duration (TZD). The parity of the mediator~\cite%
{Hutton2004,Bazhanov2018} critically determines the resulting entanglement.
Furthermore, because entanglement reduces spin purity~\cite{Stocker2022},
accurately modeling these systems requires a fully quantum treatment, as
semiclassical approaches (like the Landau-Lifshitz-Gilbert equation) fail to
capture the non-conservation of local spin magnitude~\cite%
{Mondal2021,GarciaGaitan2024}.

Enhancing the tunability of entanglement often involves enabling qubit
motion. Moving qubits have been explored in cavity quantum electrodynamics
systems, where atoms coupled to cavity photons preserve entanglement by
tuning their velocities~\cite{Mortezapour2017,Huan2015,Obada2010} or prevent
ESD by manipulating cavity-cavity interactions~\cite{Pandit2018}. In the
solid state, strong entanglement can be created by scattering ballistic
electrons off magnetic impurities~\cite{CostaJr2001,Sharma2021}. However, a
systematic and programmable method for shaping the temporal entanglement
profile---or trajectory---in solid-state systems is still lacking. Resolving
this is of fundamental and practical significance for sustained entanglement
distribution and the exploration of the dynamically accessible Hilbert space.

In this paper, we present an RKKY-exchange-based device for the systematic
shaping of entanglement trajectories. The spin qubits considered here can be
implemented using suspended impurity/defect centers or stationary
QD-confined spins coupled effectively through the mediating two-dimensional
electron gas (2DEG). We show that the dynamics are parameterized by the
exchange-time integral (ETI), which acts as an effective evolution variable
and enables the reversal of previously visited states along a trajectory. In
our implementation, the required time dependence of the sign change in the
exchange coupling is achieved by a motion-driven scheme (prescribing the
spatial vibrational motion of the qubits). Alternatively, a gate-driven
scheme, which switches the exchange between ferromagnetic and
antiferromagnetic regimes via dynamical voltages $V_{G}\left( t\right) $ 
\cite{Leon2019,Mousavi2021,Tran2024}, can realize the same exchange dynamics
without motion. We demonstrate that distinct trajectories can be achieved by
employing the boundary-proximal initial states (ISs) identified through our
recent work in Ref.~\onlinecite{Chen2025}, which are valuable for quantum
metrology due to their sensitivity to external fields~\cite{Trenyi2024}.
Furthermore, introducing damping to the out-of-phase vibrations facilitates
the generation and stabilization of large entanglement. The proposed device
enables reversible access to quantum states and supports applications
exploiting near-boundary physics, characterized by weak entanglement.
Moreover, by alternating the exchange polarity, our platform also offers an
echo-like correction mechanism~\cite{Hahn1950,Bluhm2011}, unwinding phase
accumulation to mitigate dephasing.

The paper is organized as follows. Section \ref{sec:mf} introduces the
model, describes the device, and outlines the ISs required to realize ESD,
ESB, and TFD. The subsequent numerical analysis is divided into two
distinguishable dynamic regimes based on the boundedness of the
exchange-time-integral (ETI). Section \ref{sec:inp} demonstrates cyclic
navigation via in-phase and antiphase vibrations, where the system
periodically retraces its path. In contrast, Section \ref{sec:outp} examines
stabilized navigation, showing how out-of-phase modulation accesses higher
entanglement subspaces while damping locks the system to a stationary
entanglement value. Finally, Sec. \ref{sec:summ} summarizes our findings.

\section{Model and formalism}

\label{sec:mf} In this section, we delineate the device architecture,
establish the theoretical framework for the system, and introduce the ETI as
the governing control variable. We then define the specific
boundary-proximal ISs used to investigate the entanglement dynamics. 
\begin{figure*}[tbph]
\centerline{\psfig{file=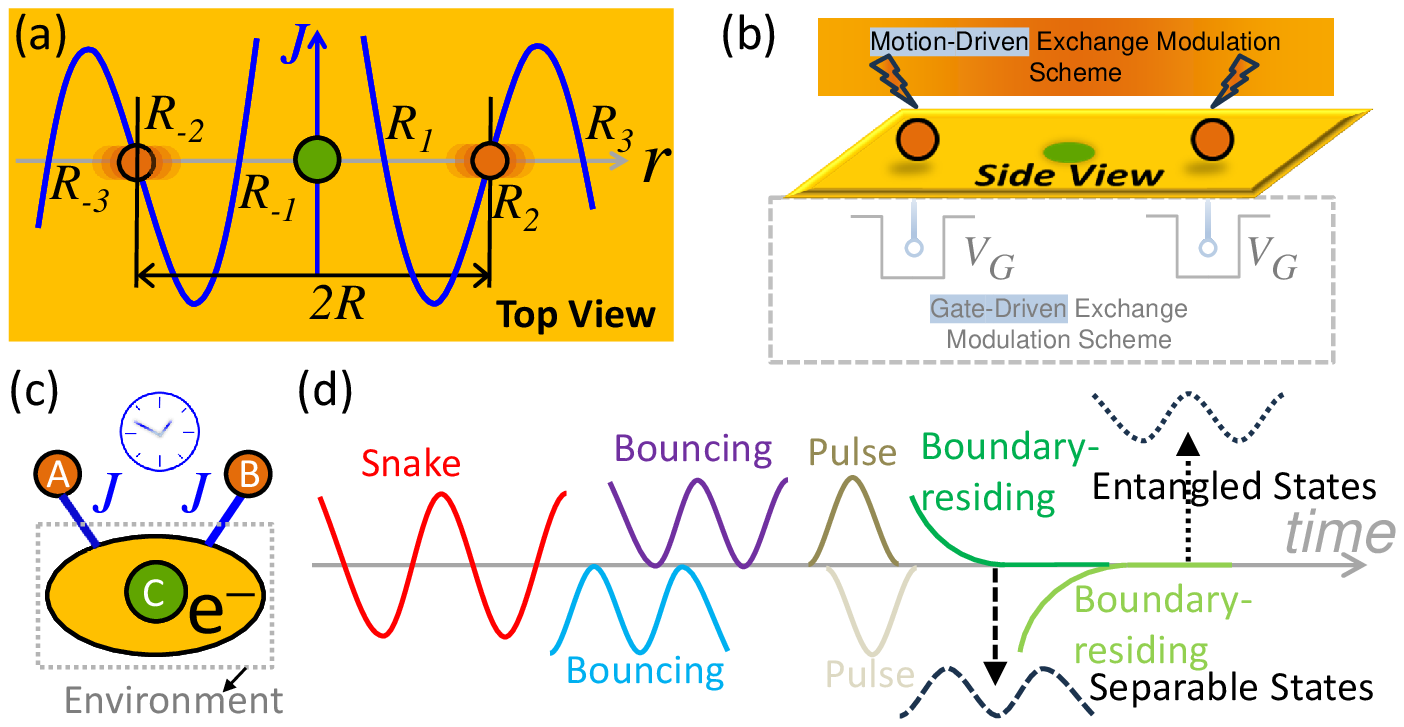,scale=0.65,angle=90,clip=true,
trim=75mm 25mm 0mm 0mm}}
\caption{System schematic and entanglement dynamics. (a) Top view of
suspended vibrating spin qubits $A$ and $B$ (orange) coupled via an
effective, alternating RKKY exchange $J$ to a central spin qudit $C$
(green). This coupling is mediated by the interaction between the qubits and
the local electron spin polarization induced by the central qudit. The
environment comprises $C$ and a two-dimensional electron gas $e$ (2DEG,
yellow), and the spatial RKKY profile (blue) defines the exchange nodes $%
R_{n}$. (b) Side view illustrating the motion-driven exchange modulation
scheme. Here, the exchange coupling to the underlying local electron spins
is modulated by the distance between the qubits and the $e$-spins
(non-contact proximity). Alternatively, a gate-driven scheme (gray dashed
box), using an applied dynamical voltage $V_{G}=V_{G}\left( t\right) $, can
generate an equivalent sign-alternating exchange interaction; in this
scheme, stationary spins are confined within the $V_{G}$-defined quantum
dots. (c) Conceptual illustration of the environment (gray dotted box) and
the qubits. The Exchange-Time-Integral (ETI)---combining the exchange
modulation (blue lines) and time (blue clock icon)---serves as the effective
control parameter that determines the qubit evolution along a trajectory in
Hilbert space. (d) Schematic representation of entanglement trajectories,
where the horizontal gray line denotes the boundary separating the entangled
(upper) and separable (lower) subspaces, highlighting distinct dynamical
regimes including snake, bouncing, pulse, and boundary-residing
trajectories. }
\label{fig:sche}
\end{figure*}

\subsection{Hamiltonian and dynamics}

\label{sec:hamiltonian}

As illustrated in Fig.~\ref{fig:sche}(a), the proposed device consists of
two spin qubits, $A$ and $B$, separated by a distance $2R$, and a mediating
environment. This environment comprises a central spin qudit $C$ (a $d$%
-level spin-$\vec{S}^{C}$ quantum system with spin $\vec{S}^{C}$) and the
itinerant electrons of a 2DEG, denoted as $e$. The central spin polarizes
the 2DEG, generating a spin-density imbalance whose sign oscillates with
distance---a hallmark of the RKKY interaction. We examine the entanglement
between the two qubits mediated by this exchange. Coupling to the
environment can be realized via two schemes: (\textit{i}) suspended qubits
interacting with local electron spins in the 2DEG via proximity effects
[Fig. \ref{fig:sche}(a)], or (ii) stationary qubits confined by QDs in the
2DEG [Fig. \ref{fig:sche}(b)]. Since both schemes generate equivalent
sign-alternating exchange (the latter via dynamical gate voltages~\cite%
{Leon2019,Mousavi2021,Tran2024}), we will focus our analysis on the case of
suspended qubits.

The local $s$-$d$ exchange interaction between the spins (qubits or qudits)
and the conduction electrons is modeled using a Dirac delta function as 
\begin{equation}
H_{sd}=J_{sd}\sum_{i\in \{A,B,C\}}\vec{S}^{i}\cdot \vec{\sigma}^{e}\,\delta (%
\vec{r}^{e}-\vec{r}^{i})\text{,}  \label{eq:Hsd}
\end{equation}%
which, upon integrating out the electron degrees of freedom, yields an
effective indirect RKKY interaction among the localized spins: 
\begin{eqnarray}
H_{\mathrm{RKKY}} &=&J(r^{AB})\,\vec{\sigma}^{A}\cdot \vec{\sigma}^{B} 
\notag \\
&&+J(r^{AC})\,\vec{\sigma}^{A}\cdot \vec{S}^{C}  \notag \\
&&+J(r^{BC})\,\vec{\sigma}^{B}\cdot \vec{S}^{C}\text{,}
\end{eqnarray}%
where $\vec{r}^{ij}=\vec{r}^{i}-\vec{r}^{j}$. Here, $\vec{r}^{A}$, $\vec{r}%
^{B}$, $\vec{r}^{C}$, and $\vec{r}^{_{e}}$ denote the position vectors of
qubits $A$ and $B$, qudit $C$, and the electron, respectively. The operators 
$\vec{\sigma}^{A}$ and $\vec{\sigma}^{B}$ represent the Pauli vector
matrices of the qubit subsystems, while $\vec{S}^{C}$ is the spin operator
of qudit $C$ (where $\vec{S}^{C}=\vec{\sigma}^{C}/2$ for the $d=2$ case).
Notably, the coupling strength scales as $J(r)\propto J_{sd}^{2}$ and, in $%
\alpha $ spatial dimensions, decays as $J(r)1\propto /r^{\alpha }$
(modulated by oscillations). Consequently, we neglect the direct coupling
between $A$ and $B$, assuming $J(r^{AB})\ll J(r^{AC})$ and $J(r^{AB})\ll
J(r^{BC})$. The qubits are subject to a local, spin-independent orbital
confinement potential $H_{o}$. The resulting spatial evolution $r^{AB}\left(
t\right) $ governed by $H_{o}$ leads to an effective spin Hamiltonian of the
form 
\begin{equation}
H=J(r^{A}\left( t\right) )\,\vec{\sigma}^{A}\cdot \!\vec{S}%
^{C}+J(r^{B}\left( t\right) )\,\vec{\sigma}^{B}\cdot \!\vec{S}^{C}\text{,}
\label{eq:Hr}
\end{equation}%
where we have set the origin at $\vec{r}^{C}\equiv 0$. The explicit time
dependence of the RKKY exchange terms, $J(r^{A}\left( t\right) )$ and $%
J(r^{B}\left( t\right) )$, originates from the motion of the qubits. This
dependence can be induced by a simple harmonic confinement potential, 
\begin{equation}
H_{o}=\frac{1}{2}k^{A}\left( \vec{r}^{A}-\vec{R}_{0}^{A}\right) ^{2}+\frac{1%
}{2}k^{B}\left( \vec{r}^{B}-\vec{R}_{0}^{B}\right) ^{2}\text{,}
\label{eq:Hshm}
\end{equation}%
which drives the vibrational motion of $A$ and $B$ about their respective
equilibrium positions, $\vec{R}_{0}^{\,A}$ and $\vec{R}_{0}^{\,B}$. However,
any mechanism that induces a dynamic sign change in $J$~\cite%
{Leon2019,Mousavi2021,Tran2024} will enable the design of desired
entanglement trajectories, such as those shown in Fig. \ref{fig:sche}(d). We
note that an alternating exchange sign can equivalently be achieved by
applying a harmonic potential to the environmental qudit $C$, rather than to 
$A$ or $B$.

The operating regime considered here is governed by the competition between
two mechanisms: the Kondo effect, which screens a local moment into a
many-body singlet~\cite{Yosida1966}, and the RKKY interaction, which
produces an oscillatory ordering of local moments~\cite%
{Cho2006,Allerdt2017,Stocker2024,Kettemann2024}. In this paper, we
concentrate on the RKKY-dominated regime, characterized on the Doniach phase
diagram by a sufficiently small exchange coupling $J_{sd}$---or,
equivalently, a low density of states at the Fermi level, $N(E_{F})$. This
condition ensures that the RKKY interaction scale, $J\sim J_{sd}^{2}N(E_{F})$%
, exceeds the Kondo energy scale $k_{B}T_{K}$, where $T_{K}\sim \exp \!\left[
-1/\left( N(E_{F})J_{sd}\right) \right] $~\cite{Doniach1977,Kroha2017}.

We focus on the motion-driven scheme, where capital $R$ denotes the
locations of the exchange nodes (zeros of $J$). Specifically, we set the
equilibrium positions $\vec{R}_{0}^{\,A}=\vec{R}_{n}$ and $\vec{R}_{0}^{\,B}=%
\vec{R}_{-n}$ such that the $A$ and $B$ qubits vibrate about the $n$-th and (%
$-n$)-th nodes, respectively, satisfying $J(\vec{R}_{0}^{\,A/B})=0$.
According to Eq. (\ref{eq:Hshm}), the spatial evolution of the qubits,
characterized by the vibrational motion parallel to the 2DEG plane, is given
by%
\begin{equation}
r^{A/B}\left( t\right) =R_{0}^{A/B}+\mathcal{R}^{A/B}\cos \left( \omega
^{A/B}t+\phi ^{A/B}\right) \text{,}  \label{eq:rAB}
\end{equation}%
with frequency $\omega ^{A/B}$, phase constant $\phi ^{A/B}$, and
displacement amplitude $\mathcal{R}^{A/B}$. We assume small amplitudes $%
\mathcal{R}^{A/B}\ll \left\vert \vec{R}_{n+1}-\vec{R}_{n}\right\vert $ so
that the exchange interaction in Eq. (\ref{eq:Hr}) can be linearized with
respect to the distance $r^{A/B}$ as%
\begin{eqnarray}
J(r^{A/B}\left( t\right) ) &\approx &J\left( R_{0}^{\,A/B}\right) +  \notag
\\
&&\left. \frac{dJ\left( r^{A/B}\right) }{dr^{A/B}}\right\vert _{\vec{r}%
^{A/B}=\vec{R}_{0}^{A/B}}\times  \notag \\
&&\left( r^{A/B}-R_{0}^{A/B}\right) \text{.}  \label{eq:Jrtlin}
\end{eqnarray}%
Substituting Eq. (\ref{eq:rAB}) into this expansion yields%
\begin{equation}
J^{A/B}(t)=J_{0}^{A/B}\cos \left( \omega ^{A/B}t+\phi ^{A/B}\right) \text{,}
\label{eq:JABt}
\end{equation}%
with the alternating-exchange amplitude defined as%
\begin{equation}
J_{0}^{A/B}=\left. \frac{dJ\left( r^{A/B}\right) }{dr^{A/B}}\right\vert
_{r^{A/B}=\vec{R}_{0}^{A/B}}\mathcal{R}^{A/B}\text{.}
\end{equation}%
Consequently, the $A$-$B$ system in Eq.~(\ref{eq:Hr}) effectively evolves
under the time-dependent Hamiltonian,%
\begin{eqnarray}
H\left( t\right) &=&J^{A}\left( t\right) \,\vec{\sigma}^{A}\cdot \!\vec{S}%
^{C}+J^{B}\left( t\right) \,\vec{\sigma}^{B}\cdot \!\vec{S}^{C}
\label{eq:HoJt1} \\
&=&J_{0}^{A}\cos \left( \omega ^{A}t+\phi ^{A}\right) \,\vec{\sigma}%
^{A}\cdot \!\vec{S}^{C}  \notag \\
&&+J_{0}^{B}\cos \left( \omega ^{B}t+\phi ^{B}\right) \,\vec{\sigma}%
^{B}\cdot \!\vec{S}^{C}\text{.}  \label{eq:HoJt}
\end{eqnarray}

\begin{table*}[tbph]
\caption{Bell-state weightings ($W_{1}\text{--}W_{14}$) and characteristic
times $T^{\ast }$ ($\hbar /J_{0}$) for mixed and pure states with $\Delta 
\protect\phi =0$. Depending on the initial weighting, the dynamics near $%
t\approx 0$ exhibit entanglement sudden death (ESD), sudden birth (ESB), or
transition of zero duration (TZD). The \emph{penetrable} entanglement switch
parameter $\protect\varepsilon $ sets whether the system begins in the
entangled regime ($\protect\varepsilon >0$, as in ESD), separable regime ($%
\protect\varepsilon <0$, as in ESB), or on the boundary ($\protect%
\varepsilon =0$).}
\label{tab:Tstar}\centering
\begin{tabular}{c|c|c|c|c}
\toprule$%
\begin{array}{c}
\text{Weighting} \\ 
\left( w_{\alpha ^{+},}w_{\alpha ^{-},}w_{\beta ^{+},}w_{\beta ^{-}}\right)%
\end{array}%
$ & $%
\begin{array}{c}
T^{\ast }\text{ (}\hbar /J_{0}\text{) for} \\ 
\text{Mixed States}%
\end{array}%
$ & $%
\begin{array}{c}
\text{Dynamics of} \\ 
\text{Mixed States Near }t\approx 0%
\end{array}%
$ & $%
\begin{array}{c}
T^{\ast }\text{ (}\hbar /J_{0}\text{) for} \\ 
\text{Pure States}%
\end{array}%
$ & $%
\begin{array}{c}
\text{Dynamics of} \\ 
\text{Pure States Near }t\approx 0%
\end{array}%
$ \\ \hline
$W_{1}=\left( \frac{1+\varepsilon }{2},\frac{1-\varepsilon }{2},0,0\right) $
& $0.6285$ & ESD & N/A & \text{TZD} \\ \hline
$W_{2}=\left( \frac{1+\varepsilon }{2},0,\frac{1-\varepsilon }{2},0\right) $
& $2.6185$ & ESD & N/A & \text{TZD} \\ \hline
$W_{3}=\left( \frac{1+\varepsilon }{2},0,0,\frac{1-\varepsilon }{2}\right) $
& $0.6283$ & ESD & N/A & \text{TZD} \\ \hline
$W_{4}=\left( 0,\frac{1+\varepsilon }{2},\frac{1-\varepsilon }{2},0\right) $
& $2.6185$ & ESD & N/A & \text{TZD} \\ \hline
$W_{5}=\left( 0,\frac{1+\varepsilon }{2},0,\frac{1-\varepsilon }{2}\right) $
& $0.6283$ & ESD & N/A & \text{TZD} \\ \hline
$W_{6}=\left( 0,0,\frac{1+\varepsilon }{2},\frac{1-\varepsilon }{2}\right) $
& N/A & \text{TZD} & N/A & \text{TZD} \\ \hline
$W_{7}=\left( \frac{1+\varepsilon }{2},\frac{1-\varepsilon }{4},\frac{%
1-\varepsilon }{4},0\right) $ & $0.8801$ & ESD & $0.5525$ & ESB \\ \hline
$W_{8}=\left( 0,\frac{1+\varepsilon }{2},\frac{1-\varepsilon }{4},\frac{%
1-\varepsilon }{4}\right) $ & $0.8779$ & ESD & $0.2822$ & ESB \\ \hline
$W_{9}=\left( \frac{1-\varepsilon }{4},0,\frac{1+\varepsilon }{2},\frac{%
1-\varepsilon }{4}\right) $ & $0.5139$ & ESD & $0.3124$ & ESD \\ \hline
$W_{10}=\left( \frac{1-\varepsilon }{4},\frac{1-\varepsilon }{4},0,\frac{%
1+\varepsilon }{2}\right) $ & $2.8819$ & ESB & $0.4422$ & ESB \\ \hline
$W_{11}=\left( \frac{1+\varepsilon }{2},\frac{1-\varepsilon }{6},\frac{%
1-\varepsilon }{6},\frac{1-\varepsilon }{6}\right) $ & $0.7652$ & ESD & $%
0.5929$ & ESB \\ \hline
$W_{12}=\left( \frac{1-\varepsilon }{6},\frac{1+\varepsilon }{2},\frac{%
1-\varepsilon }{6},\frac{1-\varepsilon }{6}\right) $ & $0.7652$ & ESD & $%
0.6036$ & ESD \\ \hline
$W_{13}=\left( \frac{1-\varepsilon }{6},\frac{1-\varepsilon }{6},\frac{%
1+\varepsilon }{2},\frac{1-\varepsilon }{6}\right) $ & $0.5444$ & ESD & $%
0.1996$ & ESD \\ \hline
$W_{14}=\left( \frac{1-\varepsilon }{6},\frac{1-\varepsilon }{6},\frac{%
1-\varepsilon }{6},\frac{1+\varepsilon }{2}\right) $ & $2.2990$ & ESB & $%
0.4406$ & ESB \\ 
\bottomrule
\end{tabular}%
\end{table*}

We elucidate the role of the dynamical exchange $J\left( t\right) $ when the
two qubits vibrate at the same frequency $\omega ^{A}=\omega ^{B}=\omega $,
specifically for the synchronous regimes of in-phase ($\Delta \phi \equiv
\phi ^{A}-\phi ^{B}=0$) and antiphase ($\Delta \phi =\pm \pi $) motion. The
system dynamics are governed by the Liouville-von Neumann equation,%
\begin{equation}
\frac{d\varrho \left( t\right) }{dt}=-i\left[ H\left( t\right) ,\varrho
\left( t\right) \right] \text{,}  \label{eq:dvarrhodt}
\end{equation}
where $\hbar \equiv 1$. The formal solution is 
\begin{equation}
\varrho \left( t\right) =U\left( t\right) \varrho _{0}U^{\dagger }\left(
t\right) \text{,}  \label{eq:varrho_t}
\end{equation}%
with the evolution operator 
\begin{equation}
U\left( t\right) =\tau \left\{ \exp \left[ -i\int\nolimits_{0}^{t}H\left(
t^{\prime }\right) dt^{\prime }\right] \right\} \text{,}  \label{eq:Ut_tau}
\end{equation}%
and the initial density matrix (DM) $\varrho \left( t=0\right) \equiv
\varrho _{0}$. Crucially, in these synchronous regimes, the time dependence
in Eq. (\ref{eq:HoJt}) becomes a common scalar factor, allowing the
Hamiltonian to be factorized into a time-dependent amplitude and a
time-independent operator. Consequently, the Hamiltonian commutes with
itself at different times ($\left[ H\left( t\right) ,H\left( t^{\prime
}\right) \right] =0$), allowing the time-ordering operator $\tau $ to be
dropped.

Consider the in-phase case where $\phi _{A}=\phi _{B}\equiv \phi $. Direct
integration yields%
\begin{eqnarray}
U\left( t\right) &=&\exp [-iI\left( t\right)  \notag \\
&&\times \sum_{k=x,y,z}\left( \sigma _{k}^{A}\eta _{J,k}^{A}+\sigma
_{k}^{B}\eta _{J,k}^{B}\right) S_{k}^{C}]  \label{eq:Ut}
\end{eqnarray}%
where the time-dependence is fully captured by the quantity%
\begin{equation}
I\left( t\right) =\frac{J_{0}}{\omega }\left[ \sin \left( \omega t+\phi
\right) -\sin \left( \phi \right) \right] \text{.}  \label{eq:It}
\end{equation}%
We refer to $I\left( t\right) $ as the ETI, which quantifies the accumulated
exchange interaction~\cite{Lin2025}. Here, we have generalized to
anisotropic exchange by defining the vector amplitudes 
\begin{equation}
\vec{J}_{0}^{A/B}=\left( J_{0,x}^{A/B},J_{0,y}^{A/B},J_{0,z}^{A/B}\right)
=J_{0}\vec{\eta}_{J}^{A/B}\text{.}  \label{eq:J0ABvec}
\end{equation}%
We set the scaling such that $J_{0}\equiv \left\vert \vec{J}%
_{0}^{A}\right\vert $ and $\vec{\eta}_{J}^{A/B}$ is a unit vector. The
coupling ratio is defined as $\gamma \equiv J_{0}^{A}/J_{0}^{B}$, implying
the relative vector components are related by 
\begin{equation}
\eta _{J,k}^{B}=\frac{\eta _{J,k}^{A}}{\gamma }\text{,}
\end{equation}%
with $k\in \left\{ x,y,z\right\} $. For antiphase motion, the condition $%
\phi ^{A}=\phi ^{B}\pm \pi $ yields a negative coupling ratio ($\gamma <0$),
as $\cos \left( \phi +\pi \right) =-\cos (\phi )$. Employing the
eigen-decomposition of the time-independent operator, 
\begin{equation}
\sum_{k=x,y,z}\left( \sigma _{k}^{A}\eta _{J,k}^{A}+\sigma _{k}^{B}\eta
_{J,k}^{B}\right) S_{k}^{C}=VDV^{\dagger }\text{,}  \label{eq:VDVdag}
\end{equation}%
the DM evolution reduces to%
\begin{eqnarray}
\varrho \left( t\right) &=&V\exp \left[ -iI\left( t\right) D\right]
V^{\dagger }  \notag \\
&&\times \varrho _{0}^{{}}  \notag \\
&&\times V\exp \left[ iI\left( t\right) D\right] V^{\dagger }\text{,}
\label{eq:varrhot}
\end{eqnarray}%
with $V$ ($D$) the eigenvector (diagonal-eigenvalue) matrix. This result
highlights that the exchange strength $J_{0}$ and time $t$ influence the
system solely through the single parameter $I\left( t\right) $. Time
reversal is thus effected by reversing the sign of the exchange profile to
retrace $I\left( t\right) $. The validity of this scalar ETI extends beyond
the linearized harmonic motion shown above. Generally, as long as the ratio
between the couplings remains constant in time---i.e., $\vec{J}^{A}\left(
t\right) =\gamma \vec{J}^{B}\left( t\right) $---the Hamiltonian factorizes
into a common time-dependent scalar factor and a static spin operator. In
such cases, the ETI takes the general form, up to a constant,%
\begin{equation}
I\left( t\right) =\int\nolimits_{0}^{t}J^{A/B}\left( t^{\prime }\right)
dt^{\prime }\text{.}  \label{eq:It_gen}
\end{equation}%
This factorization allows the complex dynamics to be mapped onto the single
ETI, enabling the designable trajectories discussed below. Conversely, the
out-of-phase case requires extending the ETI to a vector form, yielding
irregular trajectories where reversing the exchange profile sign does not
effect time reversal. This regime is detailed in Section \ref{sec:outp}.
Physically, the calculations above assume the fast-motion limit, implying
the qubit vibration frequency is sufficiently high that the spatial RKKY
profile (determined by the electron spin polarization in the 2DEG) does not
fundamentally reconfigure within one period. We further note that the
constant exchange ratio $\gamma $ required for this Hamiltonian
factorization can be engineered via coherent control of external fields,
such as the dynamical gate voltage $V_{G}\left( t\right) $ [see Fig. \ref%
{fig:sche}(b)] or applied electromagnetic waves.

To quantify the deviation of a state from the $A$-$B$
entanglement-separability boundary, we compute the reduced DM%
\begin{eqnarray}
\rho \left( t\right) &\equiv &\rho ^{AB}\left( t\right)  \notag \\
&=&Tr_{C}\left[ \varrho \left( t\right) \right] \text{,}  \label{eq:rhoABt}
\end{eqnarray}%
obtained by tracing out the spin degrees of freedom of qubit $C$. We then
employ the concurrence $\mathcal{C}_{E}$~\cite%
{Hill1997,Wootters1998,Rungta2001}, \textit{extended to admit negative values%
}, 
\begin{equation}
\mathcal{C}_{E}(t)=2\kappa _{\max }-K  \label{eq:C_E}
\end{equation}%
where $\kappa \in \left\{ \kappa _{1},\kappa _{2},\kappa _{3},\kappa
_{4}\right\} $ are the eigenvalues of the matrix 
\begin{equation}
\sqrt{\sqrt{\rho \left( t\right) }\rho ^{\prime }\left( t\right) \sqrt{\rho
\left( t\right) }}\text{,}  \label{eq:sqrtR}
\end{equation}%
with $\kappa _{\max }${}$=\max \left( \kappa \right) $, and $K=\left( \kappa
_{1}+\kappa _{2}+\kappa _{3}+\kappa _{4}\right) $. Here, $\rho ^{\prime
}\left( t\right) $ is constructed from the complex conjugate $\rho \left(
t\right) ^{\ast }$ via the transformation, 
\begin{equation}
\rho ^{\prime }\left( t\right) =\sigma _{y}^{\otimes 2}\rho \left( t\right)
^{\ast }\sigma _{y}^{\otimes 2}\text{.}
\end{equation}%
Importantly, positive values of $\mathcal{C}_{E}(t)$ indicate entanglement,
while negative values indicate separability. A larger absolute value $%
\left\vert \mathcal{C}_{E}(t)\right\vert $ signifies a greater deviation
from the boundary. As verified by our numerical calculations, this metric $%
\mathcal{C}_{E}(t)$ aligns qualitatively with the entanglement negativity 
\cite{Chen2025}.

\subsection{Boundary-proximal initial states}

\label{sec:boundary_IS}

Although Eq. \ref{eq:varrhot} allows for an explicit expression of the
evolved $\mathcal{C}_{E}(t)$ and predictions on the existence of ESD and ESB
(see Appendix \ref{apd:E_C_ETI}) in the mixed-state case, to systematically
explore the navigation of entanglement trajectories in both mixed and pure
states, we employ the ISs characterized by the entanglement switch parameter
(ESP~\cite{Chen2025}) $\varepsilon $. The ESP is defined as penetrable if it
allows the IS to be tuned across the entanglement-separability boundary: $%
\varepsilon >0$ corresponds to entangled states, while $\varepsilon <0$
denotes separable states. By selecting ISs near $\left\vert \varepsilon
\right\vert \approx 0$, we can investigate critical dynamics such as ESD and
ESB, with the exact $\varepsilon =0$ corresponding to the states at the
boundary. The initial reduced DM, $\rho _{0}\equiv \rho \left( t=0\right) $,
is expanded in the basis of the four Bell states,%
\begin{equation}
\left\vert \alpha ^{\pm }\right\rangle =\sqrt{\frac{1}{2}}\left\vert
\uparrow ,\uparrow \right\rangle \pm \sqrt{\frac{1}{2}}\left\vert \downarrow
,\downarrow \right\rangle
\end{equation}%
and%
\begin{equation}
\left\vert \beta ^{\pm }\right\rangle =\sqrt{\frac{1}{2}}\left\vert \uparrow
,\downarrow \right\rangle \pm \sqrt{\frac{1}{2}}\left\vert \downarrow
,\uparrow \right\rangle \text{,}
\end{equation}%
yielding the form%
\begin{eqnarray}
\rho _{0} &=&w_{\alpha ^{+}}\left\vert \alpha ^{+}\right\rangle \left\langle
\alpha ^{+}\right\vert +w_{\alpha ^{-}}\left\vert \alpha ^{-}\right\rangle
\left\langle \alpha ^{-}\right\vert  \notag \\
&&+w_{\beta ^{+}}\left\vert \beta ^{+}\right\rangle \left\langle \beta
^{+}\right\vert +w_{\beta ^{-}}\left\vert \beta ^{-}\right\rangle
\left\langle \beta ^{-}\right\vert \text{,}  \label{eq:Bell_rho0}
\end{eqnarray}%
where the weighting $W=\left( w_{\alpha ^{+}},w_{\alpha ^{-}},w_{\beta
^{+}},w_{\beta ^{-}}\right) $ is normalized, $w_{\alpha ^{+}}+w_{\alpha
^{-}}+w_{\beta ^{+}}+w_{\beta ^{-}}=1$. The specific weightings considered
are listed in the first column of Table \ref{tab:Tstar}, where nonzero
weights are set near $1/2$. For mixed states (Tr$[(\varrho _{0})^{2}]<1$),
we initialize the system as $\rho _{0}\otimes \left\vert \uparrow
\right\rangle \left\langle \uparrow \right\vert $, with the central spin-1/2 
$C$ set to the spin-up state $\left\vert S^{C}\right\rangle =\left\vert
\uparrow \right\rangle $. For pure states (Tr$[(\varrho _{0})^{2}]=1$), we
construct the IS as $\varrho _{0}=\left\vert \psi _{0}\right\rangle
\left\langle \psi _{0}\right\vert $, with 
\begin{equation}
\left\vert \psi _{0}\right\rangle =\sum_{\substack{ i=\alpha ^{+},\alpha
^{-},\beta ^{+},\beta ^{-}  \\ w_{i}\neq 0}}\sqrt{w_{i}}\left\vert
i^{C}\right\rangle \otimes \left\vert i\right\rangle \text{,}
\label{eq:purePsi_0}
\end{equation}%
where the summation runs over the basis states $\left\vert i\right\rangle
\in \left\{ \left\vert \alpha ^{\pm }\right\rangle ,\left\vert \beta ^{\pm
}\right\rangle \right\} $ corresponding to the nonzero weights $w_{i}$ in
Table \ref{tab:Tstar}. The environmental states $\left\vert
i^{C}\right\rangle $ map to the spin-$z$ eigenstates $\left\vert
m\right\rangle $ of $C$ arranged in descending order ($\left\vert
m=S^{C}\right\rangle $, $\left\vert S^{C}-1\right\rangle $, $\cdots $, $%
\left\vert -S^{C}\right\rangle $). The spin $S^{C}$ is chosen to match the
number of nonzero Bell components in the expansion: $S^{C}=1/2$ for $W_{1%
\text{--}6}$ , $S^{C}=1$ (qutrit) for $W_{7\text{--}10}$, and $S^{C}=3/2$
(qudit, $d=4$) for $W_{11\text{--}14}$. Generally, \textit{ISs constructed} 
\textit{from more than two Bell states} ($W_{7\text{--}14}$) \textit{are} 
\textit{penetrable}, i.e., facilitate boundary crossings (ESD/ESB) as $%
\varepsilon $ changes sign, whereas ISs limited to two Bell states ($W_{1%
\text{--}6}$) typically exhibit ESD or TZD. For instance, all $W_{1\text{--}%
6}$ \textit{pure} states yield the TZD (see Table \ref{tab:Tstar}). As the
trivial dynamics of the TZD do not result in boundary crossings, these cases
are not examined in the following results.

\section{Results}

\label{sec:rl_d}

In the simulation results presented below, bold italic letters \textit{M}
and \textit{P} in the figures denote the cases of mixed and pure states,
respectively. Unless otherwise specified, the following defaults are adopted
in the numerical simulations. Isotropic exchange is assumed within the
configuration shown in Fig. \ref{sec:inp}(a) [Eq. (\ref{eq:rAB})], with
identical vibrational frequencies ($\omega ^{A}=\omega ^{B}\equiv \omega $)
and exchange strengths ($J_{0}^{A}=J_{0}^{B}\equiv J_{0}$), yielding $\gamma
=1$. The qubit motion commences from a position outside the exchange nodes,
with the phase at $\phi _{B}=0$. For in-phase and antiphase motions (Sec. %
\ref{sec:inp}), the dynamics are governed by Eq. (\ref{eq:varrhot}), while
Eq. (\ref{eq:Ut_tau}) describes the out-of-phase motion (Sec. \ref{sec:outp}%
). All analytical results have been verified to match the direct numerical
integration of the master equation [Eq. (\ref{eq:dvarrhodt})]. Energies are
in units of $\left\vert J_{0}\right\vert \equiv 1$, and time $t$ in units of 
$\hbar /\left\vert J_{0}\right\vert $. To observe critical boundary dynamics
(ESD, ESB, or TZD), the ESP $\left\vert \varepsilon \right\vert =0.01$ is
selected, with $J_{0}=-1$. The characteristic operating period $T^{\ast }$
is determined from the time $t^{\ast }$ required for the state to reach the
entanglement-separability boundary 
\begin{equation}
\mathcal{C}_{E}(t^{\ast })=0.  \label{eq:CE_tstar}
\end{equation}%
where the relation is given by 
\begin{equation}
t^{\ast }=T^{\ast }/4\text{,}  \label{eq:smallt_star}
\end{equation}%
The resulting values of $T^{\ast }$ (and the corresponding frequency $%
f=1/T^{\ast }$) for the selected $\varepsilon $ are listed in Table \ref%
{tab:Tstar}.

Note that the RKKY-driven spin dynamics can be disrupted by ferromagnetic
resonance (FMR) in the few-GHz range, where resonant energy absorption
enhances spin precession. To ensure that the RKKY exchange remains the
dominant mechanism, operation away from this resonance regime is required.
This can be achieved by reducing the vibrational amplitude or using a
proximate 2D Fermi-sea substrate [Fig. \ref{fig:sche}(b)] yielding $%
\left\vert J_{0}\right\vert \lesssim 1$ $\mu $eV; this upper limit
corresponds to a characteristic frequency $\left\vert J_{0}\right\vert
/\hbar \approx 1.52\times 10^{9}$ rad/s (or equivalently, $0.24$ GHz), which
is well separated from the FMR band. Furthermore, bearing in mind that sign
changes in $J^{A/B}\left( t\right) $ signify time and trajectory reversal,
we can tailor the entanglement profile as demonstrated below.

\subsection{In-phase and antiphase vibrations}

\begin{figure*}[tbph]
\centerline{\psfig{file=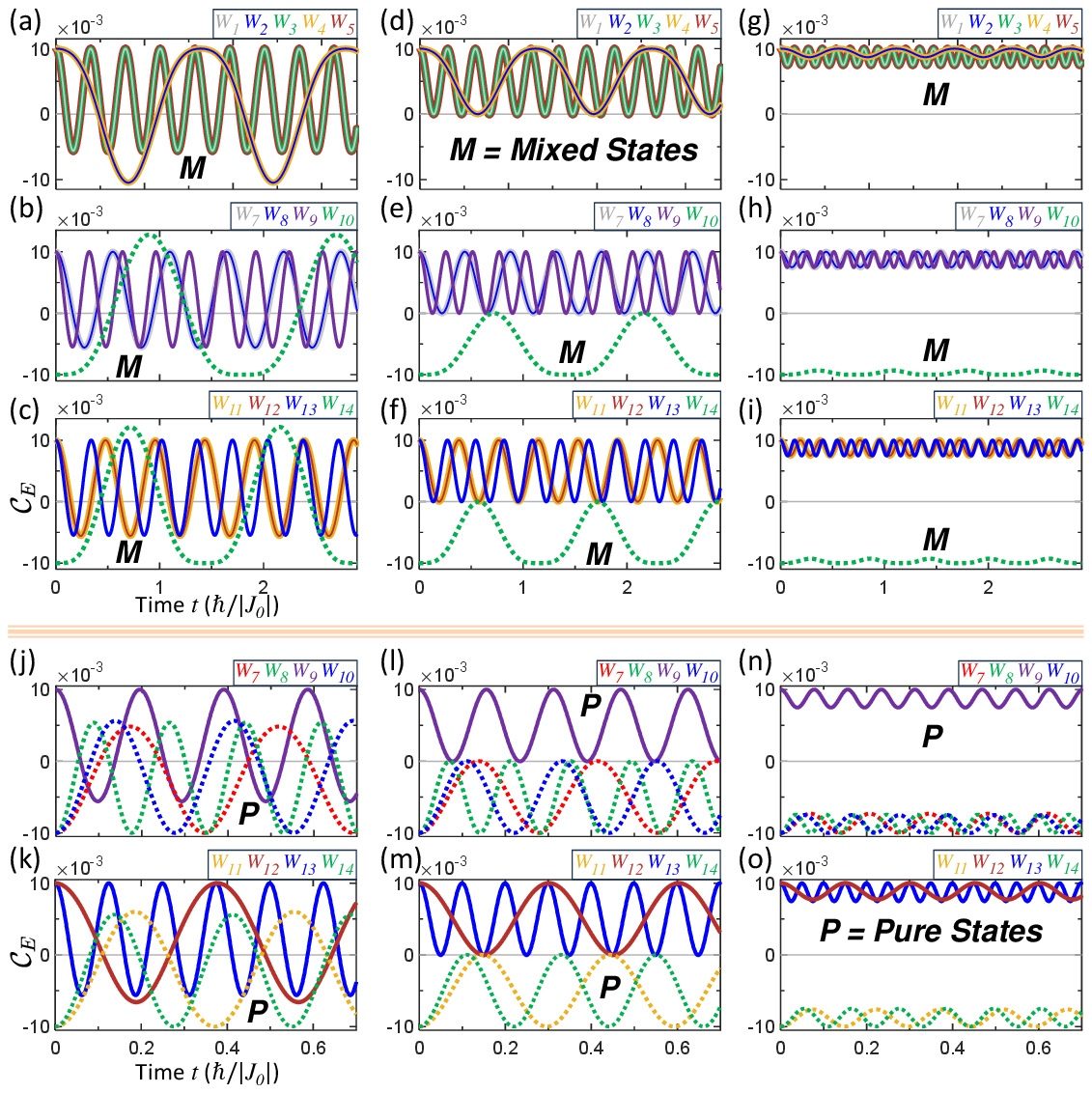,scale=0.92,angle=90,clip=true,
trim=10mm 70mm 0mm 0mm}}
\caption{Entanglement trajectories for mixed (marked by bold italic \textit{M%
}) and pure (marked by bold italic \textit{P}) states under in-phase
vibrations. The extended concurrence $\mathcal{C}_{E}(t)$ is plotted versus
time $t$ for various Bell-state weightings $W_{1\text{--}14}$ (see
corresponding color text labels). The upper panels (a)--(i) display the
dynamics for mixed states: (a)--(c) show snake trajectories ($T=1.25$ $%
T^{\ast }$), (d)--(f) show bouncing trajectories ($T=1.25$ $T^{\ast }$), and
(g)--(i) exhibit entangled/separable trajectories ($T=1.25$ $T^{\ast }$).
The lower panels (j)--(o) show the corresponding dynamics for pure states:
snake trajectories in (j) and (k), bouncing trajectories in (l) and (m), and
entangled/separable trajectories in (n) and (o). For the bouncing
trajectories, solid lines denote initially entangled states ($\protect%
\varepsilon >0$), while dashed lines denote separable states ($\protect%
\varepsilon <0$).}
\label{fig:mixpure_sbe}
\end{figure*}

\label{sec:inp} This section examines the entanglement trajectories for
in-phase and antiphase vibrations. Because the underlying analysis is
identical in both cases, as the exchange-time integral (ETI) applies
equally, we restrict the discussion to the in-phase case. For mixed states,
the characteristic period $T^{\ast }$, defined by $\mathcal{C}_{E}(T^{\ast
}/4)=0$, can be estimated analytically in the short-time regime as detailed
in Appendix \ref{apd:Ana_est_Tstar}. Substituting $T^{\ast }$ into Eq. (\ref{eq:It})
yields the weight-dependent characteristic ETI, 
\begin{equation}
I_{W}^{\ast }\equiv I\left( T^{\ast }/4\right) \text{.}
\end{equation}%
Since the entanglement dynamics depend solely on the ETI, any exchange
dynamics---whether or not it involves vibrations---described by the general
expression (\ref{eq:It_gen}) in Eq. (\ref{eq:varrhot}), with the same
initial weight $W$ and an ETI satisfying $I_{W}^{\ast }$, will necessarily
reach the entanglement--separability boundary at $t=T^{\ast }/4$. The
trajectories can therefore be systematically engineered through frequency
control. For vibrations with $T=T^{\ast }$, the ETI indicates a trajectory
reversal at $t=T^{\ast }/4$. These trajectories correspond to reversible
navigation, in which the qubit-state evolution is governed by the periodic
accumulation and unwinding of the ETI, so that reversing the sign of $%
J\left( t\right) $ effectively retraces the entanglement history.
Consequently, oscillatory trajectories arise for $T\gtrsim T^{\ast }$, $%
T=T^{\ast }$, and $T\lesssim T^{\ast }$, corresponding to snake, bouncing,
and entangled- or separable-confined profiles, respectively. The snake
trajectory (for slightly larger $T\gtrsim T^{\ast }$) periodically crosses
the entanglement-separability boundary. The bouncing trajectory (for $%
T=T^{\ast }$) reverses immediately upon reaching the boundary, corresponding
to a TZD. The entangled/separable trajectory (for slightly smaller $%
T\lesssim T^{\ast }$) remains confined to one subspace without crossing the
boundary. These trajectories are shown in Fig. \ref{fig:mixpure_sbe} for
both mixed and pure states.

\begin{figure}[tbph]
\centerline{\psfig{file=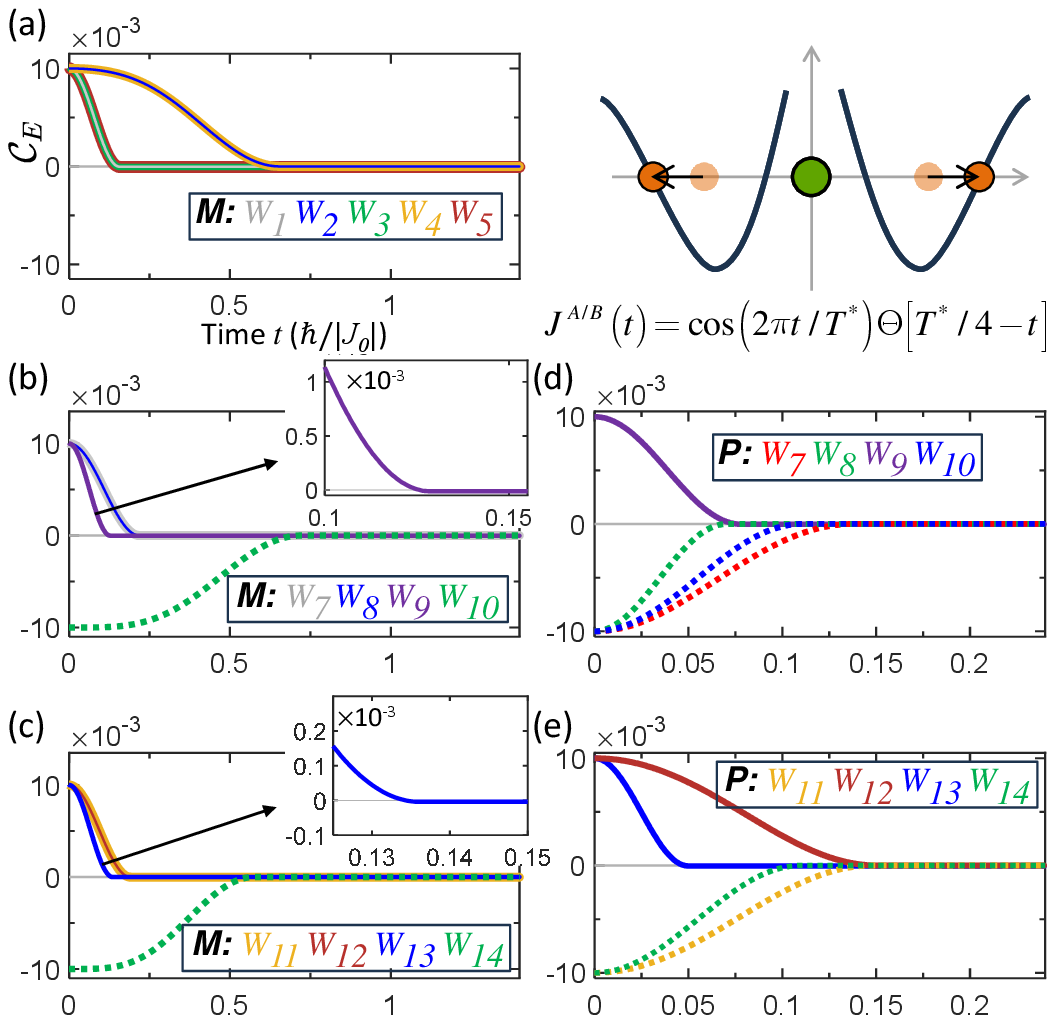,scale=0.48,angle=90,clip=true,
trim=24mm 80mm 0mm 0mm}}
\caption{Entanglement trajectories originating from abruptly halted
vibrational motion (schematic in the upper right) for (a)--(c) mixed states
and (d)--(e) pure states with different weightings. The qubits remain at the
entanglement-separability boundary after $t\geq T^{\ast }/4$, producing a
boundary-residing trajectory. The insets show zoomed views near the stopping
time $t=T^{\ast }/4$. All trajectories approach the boundary tangentially
(smoothly), despite the abrupt cessation of motion.}
\label{fig:bstay_mixpure}
\end{figure}

\begin{figure}[tbph]
\centerline{\psfig{file=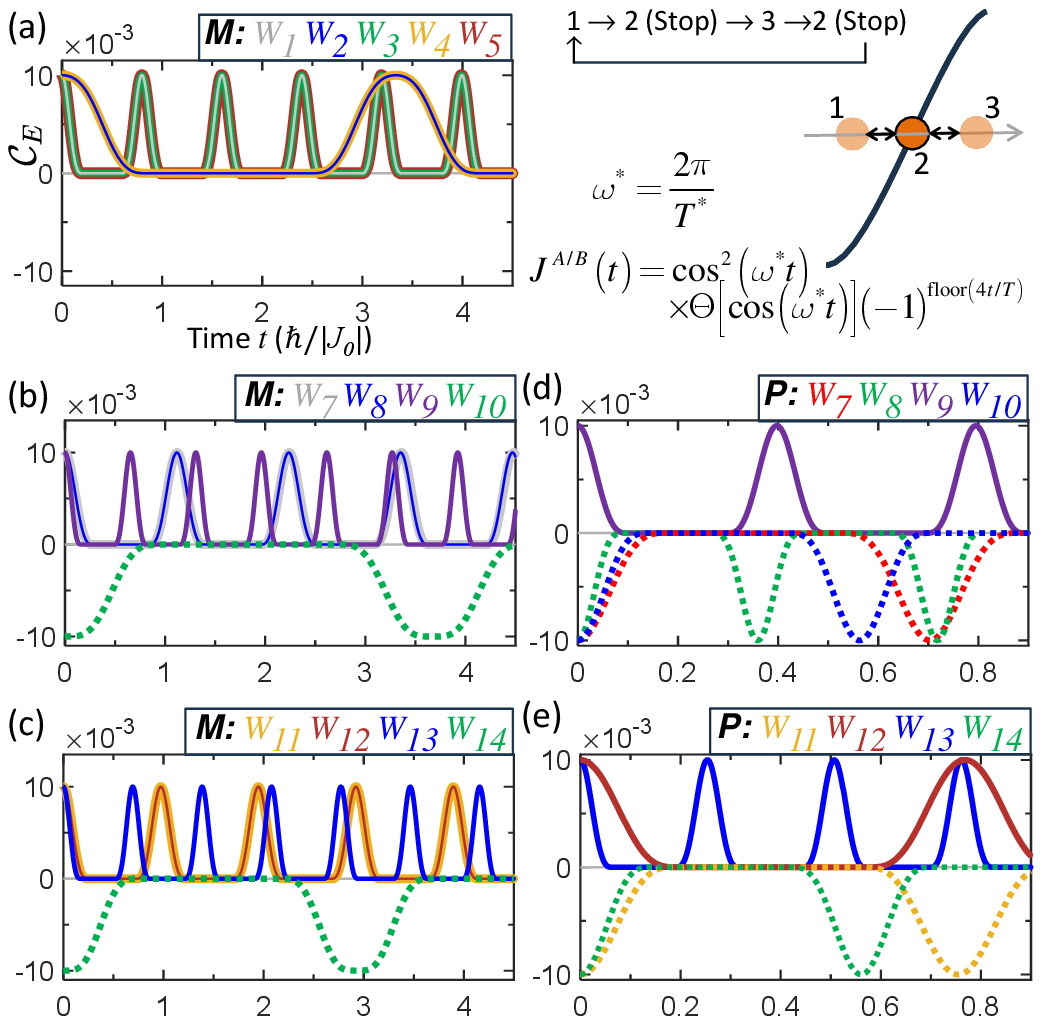,scale=0.48,angle=90,clip=true,
trim=24mm 82mm 0mm 0mm}}
\caption{Entanglement pulses generated by controlled stopping and restarting
of qubit cyclic motion, as shown schematically in the upper-right panel. The
qubits halt temporarily at the exchange nodes, producing entangled (solid)
and separable (dashed) pulses. Panels (a)--(c) show mixed states, while
panels (d) and (e) show pure states. The trajectories include repeated
boundary-residing segments, with the qubits departing from and returning to
the boundary between these segments, thereby forming pulse trains.}
\label{fig:pulse_mixpure}
\end{figure}

\begin{figure}[tbph]
\centerline{\psfig{file=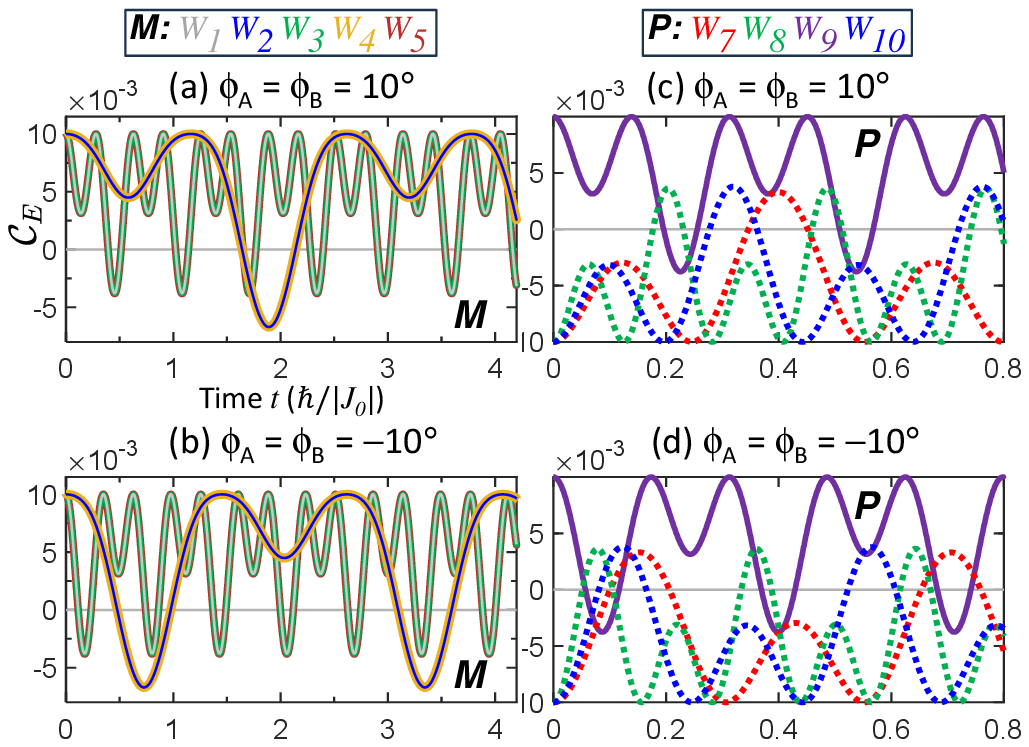,scale=0.48,angle=90,clip=true,
trim=68mm 82mm 0mm 0mm}}
\caption{Entanglement trajectories for qubits initialized with a finite
vibrational phase $\protect\phi \neq 0$. Qubits vibrating toward ($\protect%
\phi =10%
{{}^\circ}%
$) or away ($\protect\phi =-10%
{{}^\circ}%
$) from the exchange nodes with period $T=T^{\ast }$ in Table \protect\ref%
{tab:Tstar} produce snake trajectories. Panels (a) and (b) show mixed states
with weightings $W_{1\text{--}5}$, and panels (c) and (d) show pure states
with weightings $W_{7\text{--}10}$. A finite $\protect\phi $ causes
asymmetric vertical shifts, while reversing its sign introduces horizontal
phase shifts. Solid and dashed lines represent positive ($\protect%
\varepsilon >0$) and negative ($\protect\varepsilon <0$) ESP, respectively.}
\label{fig:pshift_mixpure}
\end{figure}

\begin{figure}[tbph]
\centerline{\psfig{file=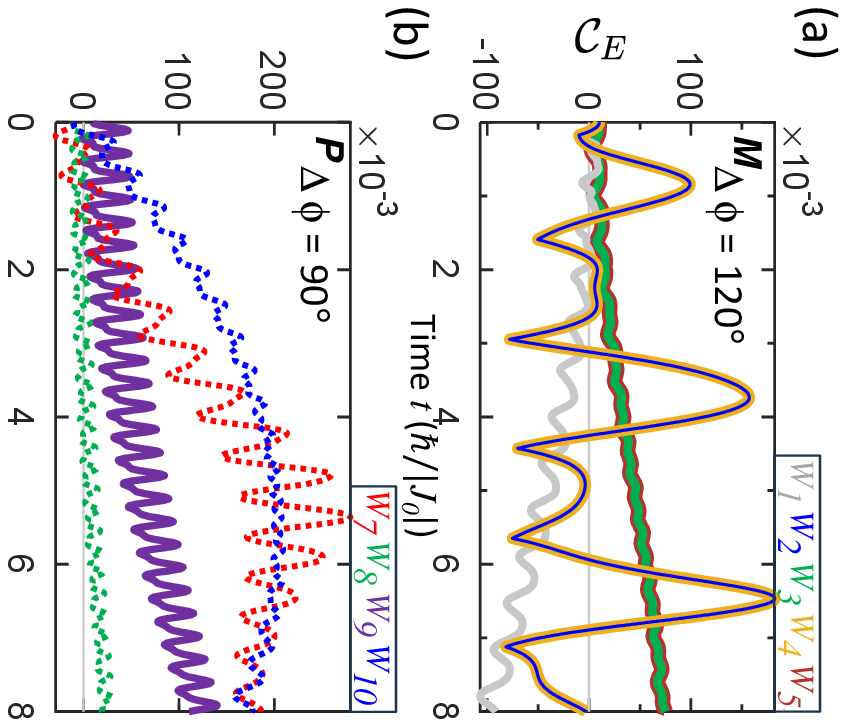,scale=0.6,angle=90,clip=true,
trim=58mm 125mm 0mm 0mm}}
\caption{Out-of-phase entanglement trajectories for (a) mixed states with $%
\Delta \protect\phi =120%
{{}^\circ}%
$ and (b) pure states with $\Delta \protect\phi =90%
{{}^\circ}%
$. Solid lines represent initially entangled states with positive ESP, while
dashed lines represent initially separable states with negative ESP.
Vibration periods $T=T^{\ast }$ from Table \protect\ref{tab:Tstar} are used.
The curves illustrate how out-of-phase motion drives the entanglement
trajectories away from the boundary (horizontal gray lines).}
\label{fig:outofp_mixpure}
\end{figure}

\begin{figure*}[tbph]
\centerline{\psfig{file=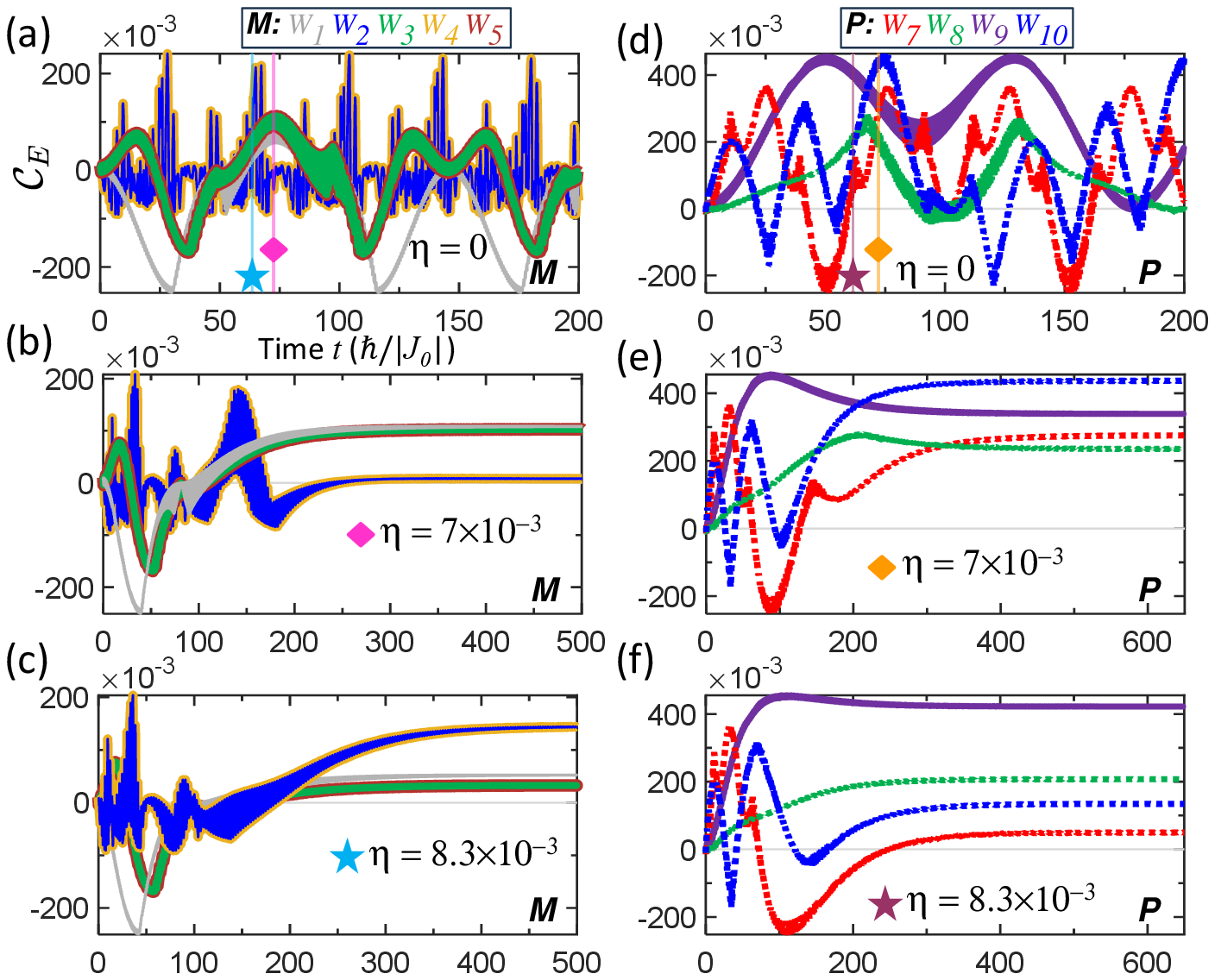,scale=0.7,angle=90,clip=true,
trim=8mm 30mm 0mm 0mm}}
\caption{Entanglement evolution for (a)--(c) mixed states and (d)--(f) pure
states with out-of-phase vibrations at $\Delta \protect\phi =30%
{{}^\circ}%
$. The vibration period $T=T^{\ast }$ from Table \protect\ref{tab:Tstar} is
used. The undamped cases ($\protect\eta =0$) are shown in (a) and (d). The
damped cases are shown with damping strength $\protect\eta =7\times 10^{-3}$
(in units of $\left\vert J_{0}\right\vert /\hbar $) in (b) and (e) (diamond
markers), and $\protect\eta =8.3\times 10^{-3}$ in (c) and (f) (star
markers). The vertical lines in (a) and (d) indicate the corresponding
stabilization time $t_{s}$ observed in the damped regimes. The transient
band-like curves arise from fast and closely spaced entanglement
modulations. A weaker damping strength leads to a larger postponement of $%
t_{s}$. Under damping, time effectively slows down, causing the trajectories
to appear stretched before $t_{s}$ and ultimately resulting in a time-frozen
trajectory. For example, $W_{7}$ in (d) shows a minimum around $t=50$, while
in (e) and (f) this minimum occurs later, around $t=100$.}
\label{fig:dampoutofp_mixpure}
\end{figure*}

For entanglement trajectories based on non-periodic motion, we consider the
same vibrational scenario, except that the qubit motion is abruptly halted
upon reaching the exchange nodes. The exchange coupling follows Eqs. (\ref%
{eq:rAB}) and (\ref{eq:JABt}), now modulated by a unit-step function $\Theta 
$,%
\begin{equation}
J^{A/B}(t)=J_{0}\cos \left( \frac{2\pi t}{T^{\ast }}\right) \Theta \left( 
\frac{T^{\ast }}{4}-t\right) \text{.}  \label{eq:JABt_cos_us}
\end{equation}%
Figure \ref{fig:bstay_mixpure} shows the resulting trajectories for both
mixed and pure states. After $t$ $\geq $ $T^{\ast }/4$, the qubits remain
pinned at the entanglement-separability boundary, forming a
boundary-residing trajectory. Notably, although the motion stops abruptly,
the entanglement exhibits a smooth approach to zero $\left. d\mathcal{C}%
_{E}(t)/dt\right\vert _{t=T^{\ast }/4}=0$ in Fig. \ref{fig:bstay_mixpure}. As a
result, neither ESD nor ESB occurs. This demonstrates that an abrupt halt of
the qubit motion, even involving a kinetic discontinuity (abrupt change in
velocity), does not necessarily induce sudden entanglement dynamics.

Alternatively, the boundary-residing behavior is also accessible by
employing a smooth (continuous-velocity) stop, where the exchange in Eq. (%
\ref{eq:JABt_cos_us}) follows $\cos ^{2}(\theta )$ instead of cos$\left(
\theta \right) $. In this case, $T^{\ast }$ is recomputed according to Eqs. (%
\ref{eq:smallt_star}) and (\ref{eq:CE_tstar}). Specifically, this protocol
realizes a pulse trajectory by employing the time dependence%
\begin{eqnarray}
J^{A/B}(t) &=&J_{0}\cos ^{2}\left( \frac{2\pi t}{T^{\ast }}\right) \Theta
\left( \cos \left( \frac{2\pi t}{T^{\ast }}\right) \right)  \notag \\
&&\times \left( -1\right) ^{floor\left( 4t/T^{\ast }\right) }\text{,}
\label{eq:JABt_cos2}
\end{eqnarray}%
which corresponds to a temporary dwell of the motion whenever the qubit
reaches an exchange node (see the schematics in Fig. \ref{fig:pulse_mixpure}%
). Here, the function $floor$ denotes rounding toward negative infinity.
Figure \ref{fig:pulse_mixpure} shows the resulting pulse trajectories for
both mixed and pure states obtained from Eq. (\ref{eq:JABt_cos2}). The dwell
duration is $T^{\ast }/2$, spanning the intervals $\left( 2p+1\right)
T^{\ast }/4$ to $\left( 2p+2\right) T^{\ast }/4$ with $p=0,1,2,\cdots $.
This sequence produces repeated boundary-residing segments, generating
entangled (solid lines) and separable (dashed lines) pulse trains. A single
pulse can be produced by imposing a permanent stop. We note, however, that,
for boundary-residing and pulsed trajectories, quantum position fluctuations 
$\delta r$ prevent the qubit position from being perfectly fixed at the
exchange node. These fluctuations can be mitigated using position-squeezed
states~\cite{Walls1983,Wu1987,Pirkkalainen2015,Marti2024}, ensuring that $%
\delta r$ remains much smaller than the vibration amplitude. In practice,
for entangled (separable) pulses, a slightly larger value of $T^{\ast
} $ is chosen so that the fluctuation in the extended concurrence $\delta 
\mathcal{C}_{e}$ remains well below (above) the boundary. Due to the
uncertainty principle, the boundary-residing trajectory is fundamentally
approximate in the motion-driven scheme, leading to residual fluctuations
around the boundary.

When qubits are initialized closer to the exchange nodes, a finite
vibrational phase $\phi \neq 0$ can be introduced. This modification
preserves the tunability of the trajectories via the period but introduces
asymmetry in the entanglement evolution. As shown in Fig. \ref%
{fig:pshift_mixpure}, using $T=T^{\ast }$ from Table \ref{tab:Tstar},
similar snake trajectories emerge for selected weightings: $W_{1\text{--}5}$
for mixed states and $W_{7\text{--}10}$ for pure states. To understand the
origin of the snake profile, we compare it to the standard ($\phi =0$)
bouncing trajectories. While the vibrational amplitude is identical, the
finite phase implies that the qubits are initialized closer to the exchange
node. Consider an entangled IS with $\phi >0$. After $t=0$, the interval
before the qubits reach the node is shortened, leading to a reduced
accumulation of the ETI and a correspondingly weaker loss of entanglement.
This initial phase therefore produces a shallow reversal entirely within the
entangled subspace. Subsequently, after passing through the node, the qubits
traverse the full range of motion until returning to the node. The ETI
accumulated during this interval matches the standard bouncing case and
exceeds the initial minor reduction. Consequently, this second swing drives
the trajectory across the boundary into the separable subspace, giving rise
to the characteristic snake profile. By the same reasoning, this snake
trajectory is also realized for separable ISs.

In other words, a finite $\phi $ induces asymmetric vertical shifts,
reflecting the difference in the ETI accumulation between consecutive
entanglement recoveries. By contrast, reversing the sign of $\phi $ means
the qubits initially move in the opposite direction; this effectively
introduces a horizontal phase shift, as observed by comparing panels (a)
with (b) and (c) with (d) in Fig. \ref{fig:pshift_mixpure}. The same
argument applies to all other weightings exhibiting ESD and ESB listed in
Table \ref{tab:Tstar}. For brevity, only the representative cases (mixed
states with $W_{1\text{--}5}$ and pure states with $W_{7\text{--}10}$) are
shown in the subsequent figures.

Importantly, for both in-phase and antiphase motion, we emphasize that the
proposed device---characterized by alternating exchange signs---possesses
built-in error correction against dephasing for oscillatory trajectories,
Figs. \ref{fig:mixpure_sbe}, \ref{fig:pulse_mixpure}, and \ref%
{fig:pshift_mixpure}. Phase coherence is preserved because the rapid phase
accumulation during the ferromagnetic ($J<0$) intervals is compensated by
the rapid phase unwinding during the antiferromagnetic ($J>0$) intervals,
refocusing the state and producing an echo-like correction~\cite%
{Hahn1950,Bluhm2011}. Notably, being exchange-free, the node serves as an
ideal location for qubits to idle. Conversely, prior to applying the
confinement potential that yields the vibrations, the stable ground state
resides at either the local maximum or minimum of $J$, making these points
natural sites for entanglement initialization and development. While
in-phase motion provides a robust baseline for tailoring entanglement
dynamics, the strict periodicity of the evolution limits the accessible
state space, particularly at short operation times. To overcome this
limitation and access larger entanglement values, we now turn to the
out-of-phase configuration, where the vibrational phases are shifted
relative to one another.

\subsection{Out-of-phase vibrations}

\label{sec:outp}

The introduction of a relative vibrational phase difference $\Delta \phi
=\phi ^{A}-\phi ^{B}\neq 0$ defines the regime of out-of-phase vibrations.
In this regime, the time-evolution operator can no longer be parameterized
by a single ETI $I\left( t\right) $, as was done in Eq. (\ref{eq:Ut}).
Instead, the dynamics are governed by two distinct integrals, $I^{A}\left(
t\right) $ and $I^{B}\left( t\right) $. Although these integrals share the
same functional form, their relative phase mismatch prevents the DM in Eq. (%
\ref{eq:varrhot}) from retaining the strict periodic structure associated
with a unified $I\left( t\right) $. Consequently, the resulting entanglement
evolution becomes generally non-periodic.

Indeed, as shown in Fig. \ref{fig:outofp_mixpure}, the concurrence ceases to
follow the simple repeating patterns observed in the in-phase or antiphase
cases. Panels (a) and (b) illustrate the evolution for mixed and pure
states, with phase differences of $\Delta \phi =120%
{{}^\circ}%
$ and $\Delta \phi =90%
{{}^\circ}%
$, respectively. The trajectories tend to drift progressively farther from
the entanglement-separability boundary, exhibiting a slow overall growth
upon which fast modulations (sub-oscillations) are superimposed. Notably,
weightings $W_{3}$ and $W_{5}$ (mixed states) and $W_{7\text{--}10}$ (pure
states) achieve significantly higher entanglement values than their $\Delta
\phi =0$ counterparts. Furthermore, $W_{2}$ and $W_{4}$ (mixed states)
exhibit swings with growing amplitude that cross the boundary multiple
times. This complex behavior arises from the interference between the
distinct accumulation rates of $I^{A}\left( t\right) $ and $I^{B}\left(
t\right) $.

Figure \ref{fig:outofp_mixpure} demonstrates that out-of-phase motion
enables access to quantum states and high-entanglement subspaces that are
inaccessible under strictly periodic motion ($\phi ^{A}=\phi ^{B}$). This
regime highlights the potential of phase-engineered, alternating RKKY
coupling to create highly versatile and non-repetitive entanglement
trajectories. However, a key limitation in this scenario is that the
trajectories do not naturally converge to a stable value, restricting their
direct applicability for state preparation. To address this, we introduce
damping mechanisms that provide a route to stabilize the trajectories.

In the out-of-phase regime where $\phi ^{A}\neq \phi ^{B}$, the exchange
integrals $I^{A}\left( t\right) $ and $I^{B}\left( t\right) $ are generally
not proportional. However, the introduction of damping fundamentally alters
this behavior. Unlike the previous cases, damping causes the vibrational
amplitude to decay, driving each ETI to converge to a distinct fixed value, $%
I^{A/B}\left( t\rightarrow \infty \right) $, as the qubits eventually settle
at the exchange node ($J=0$). As a result, the system enters a frozen
regime, stabilizing the generated entanglement. For the case of isotropic
exchange modeled by $J^{A/B}(t)=J_{0}^{{}}\cos \left( \omega ^{A/B}t+\phi
^{A/B}\right) \exp \left( -\eta \ast t\right) $, the ETI can be expressed
using Eq. (\ref{eq:It_gen}) as

\begin{eqnarray}
I^{A/B}\left( t\right) &=&\frac{-J_{0}}{\eta ^{2}+\omega ^{2}}F^{A/B}\left(
\omega ,\eta ,t\right) \exp \left( -\eta \ast t\right)  \notag \\
&&+\frac{J_{0}}{\eta ^{2}+\omega ^{2}}F^{A/B}\left( \omega ,\eta ,0\right) 
\text{,}  \label{eq:IABt_long}
\end{eqnarray}%
where the function $F^{A/B}\left( \omega ,\eta ,t\right) $ is defined as%
\begin{eqnarray}
F^{A/B}\left( \omega ,\eta ,t\right) &=&\eta \cos (\omega t+\phi ^{A/B}) 
\notag \\
&&-\omega \sin \left( \omega t+\phi ^{A/B}\right)  \notag \\
&=&\sqrt{\eta ^{2}+\omega ^{2}}  \notag \\
&&\times \cos \left[ \omega t+\phi ^{A/B}+\arctan \left( \frac{\omega }{\eta 
}\right) \right] \text{.}
\end{eqnarray}%
Here, $\eta $ represents the damping strength in units of $\left\vert
J_{0}\right\vert /\hbar $. To streamline the analysis, we introduce the
amplitude scaling factor,%
\begin{equation}
\Gamma \equiv \frac{J_{0}}{\sqrt{\eta ^{2}+\omega ^{2}}}\text{,}
\end{equation}%
and the shifted phase,%
\begin{equation}
\Phi ^{A/B}\equiv \phi ^{A/B}+\arctan \left( \frac{\omega }{\eta }\right) 
\text{.}
\end{equation}%
With these definitions, Eq. (\ref{eq:IABt_long}) simplifies to%
\begin{eqnarray}
I^{A/B}\left( t\right) &=&-\Gamma \cos \left( \omega t+\Phi ^{A/B}\right)
\exp \left( -\eta \ast t\right)  \notag \\
&&+I^{A/B}\left( t\rightarrow \infty \right) \text{,}  \label{eq:IABt_mid}
\end{eqnarray}%
where the ETI asymptotically approaches the constant stationary value,%
\begin{equation}
I^{A/B}\left( \infty \right) =\Gamma \cos \left( \Phi ^{A/B}\right) \text{.}
\label{eq:IABtinfity}
\end{equation}%
Since the cosine function is bounded by unity (i.e., $-1\leq \cos \left(
\omega t+\Phi \right) \leq 1$), the ETI satisfies the inequality,%
\begin{eqnarray}
&&\Gamma \cos \left( \Phi ^{A/B}\right) -\left\vert \Gamma \right\vert \exp
\left( -\eta \ast t\right)  \notag \\
&\leq &I^{A/B}\left( t\right)  \notag \\
&\leq &\Gamma \cos \left( \Phi ^{A/B}\right) +\left\vert \Gamma \right\vert
\exp \left( -\eta \ast t\right) \text{.}  \label{eq:IABt_bound}
\end{eqnarray}%
This bound offers valuable physical insights. First, it shows that weaker
damping (smaller $\eta $) and lower vibration frequencies (smaller $\omega $%
) expand the accessible range of the ETI by increasing the magnitude of $%
\Gamma $. Accordingly, this expansion allows for greater achievable
entanglement. Second, by comparing Eq. (\ref{eq:IABt_bound}) with Eq. (\ref%
{eq:IABtinfity}), we observe that $I^{A/B}\left( t\right) $ can transiently
exceed its stationary value $I^{A/B}\left( \infty \right) $, particularly
when $\eta $ is small. Accordingly, the transient entanglement $\mathcal{C}%
_{E}(t)$ can surpass the final stabilized value $\mathcal{C}_{E}(\infty )$.
Thus, by tuning $\omega $ and $\eta $, one can engineer the dynamics to
target a specific, stable value of entanglement.

The features of this damped evolution are illustrated in Fig. \ref%
{fig:dampoutofp_mixpure} for a representative phase difference of $\Delta
\phi =30%
{{}^\circ}%
$, where panels (a)--(c) and (d)--(f) display results for mixed and pure
states, respectively. Panels (a) and (d) present the undamped reference
cases, while the subsequent panels introduce damping with $\eta =7\times
10^{-3}$ (diamond markers) and slightly stronger damping with $\eta
=8.3\times 10^{-3}$ (star markers). Clearly, decreasing the damping strength
delays the stabilization time $t_{s}$---the characteristic time scale beyond
which the entanglement becomes effectively fixed. This delay is evident when
comparing the damped panels (b) with (c) and (e) with (f), with the vertical
lines in panels (a) and (d) marking the corresponding onset $t_{s}$ of the
frozen regime observed in the damped cases. In the early dynamics ($t<t_{s}$%
), the rapid entanglement modulations are so dense that they appear as
continuous bands. During this phase, the trajectory profile retains the
overall shape of the undamped case but with reduced amplitude and a slower
effective evolution. However, the damping also causes the evolution to slow
down, effectively stretching the profile along the time axis. For example,
the first local minimum of $W_{3}$ in the undamped motion (a) occurs around $%
t=30$, whereas in the damped motion (b) and (c), the corresponding first
local minimum appears later, after $t=45$. This delay reflects the slower
accumulation of both $I^{A}\left( t\right) $ and $I^{B}\left( t\right) $,
and thus the evolution, which eventually culminates in a time-frozen
trajectory.

Accordingly, by optimizing the damping strength, one can maximize the final
stable entanglement. As shown in Fig. \ref{fig:dampoutofp_mixpure}, specific
weightings---$W_{1}$, $W_{3}$ , $W_{5}$ in (b); $W_{2}$, $W_{4}$ in (c); $%
W_{7}$, $W_{8}$, $W_{10}$ in (e); and $W_{9}$ in (f)---yield frozen
entanglement values that significantly exceed those achievable in the $%
\Delta \phi =0$ case. We note that $I^{A/B}\left( t\right) $ in Eq. (\ref%
{eq:IABt_mid}) reduces to Eq. (\ref{eq:It}) only in the limit $\eta =0$ and $%
\Delta \phi =0$. In the damped, out-of-phase scenario, $I^{A}\left( t\right) 
$ is no longer proportional to $I^{B}\left( t\right) $, marking a
fundamental departure from the single-ETI eigen-decomposition used in Eq. (%
\ref{eq:varrhot}). Physically, this damping model is also realizable in a
scenario where the two qubits, $A$ and $B$, move in opposite directions away
from the central qudit $C$, thereby experiencing a spatially decaying,
alternating RKKY interaction. \emph{Up to here...}

\section{Conclusions}

\label{sec:summ}

In conclusion, to navigate qubit entanglement trajectories, we propose an
RKKY-based platform, Fig. \ref{fig:sche}, in which two spin qubits, $A$ and $%
B$, couple to a central spin qudit $C$ that induces an oscillatory spin
polarization in the surrounding conduction electrons. To quantify the
deviation from the entanglement-separability boundary, the concurrence is
extended to include negative values; this extension enables a unified
description of boundary-crossing dynamics. For qubits experiencing
alternating exchange $J$ around the node, the linearization (\ref{eq:Jrtlin}%
) captures the relevant dynamics, yielding an effective time-dependent
Hamiltonian (\ref{eq:HoJt1}). The qubit spatial motion is mapped onto the
ETI through the time dependence of the exchange interaction. The ETI (\ref%
{eq:It_gen}), obtained by integrating the time-dependent exchange, acts as a
single experimentally accessible control variable that parameterizes the
state evolution via (\ref{eq:varrhot}). With our interest placed on the
spin-qubit entanglement rather than on the environmental qudit and
conduction electrons, the orbital or spatial degrees of freedom are traced
out, preserving the form of the ETI-ruled evolution for any spin-independent 
$H_{o}$.

We identify two complementary dynamical regimes, cyclic and non-cyclic
navigation. First, cyclic navigation corresponds to periodic motion,
enabling the recurrent restoration of the state. Specifically, for in-phase
and anti-phase vibrations under a harmonic confinement potential, the ETI
reduces to (\ref{eq:It}). Using weighted Bell-state DMs, Eq. (\ref%
{eq:Bell_rho0}), we demonstrate frequency-controlled transitions between
snake, bouncing, and boundary-residing trajectories for both mixed states
and pure states (Figs. \ref{fig:mixpure_sbe} and \ref{fig:bstay_mixpure}).
The FMR can be avoided by selecting realistic exchange values that yield the
operating frequencies listed in Table \ref{tab:Tstar}; the corresponding ETI
values are computed accordingly (see Table \ref{tab:est_Tstar} in Appendix~\ref{apd:E_C_ETI}). By allowing temporary
dwells at the nodes, entanglement pulse trains can be generated, with pulse
separation controlled by the dwell duration, as shown in Fig. \ref%
{fig:pulse_mixpure}. A nonzero equal vibrational phase, $\phi =\phi
^{A}=\phi ^{B}$, induces an asymmetric vertical shift, while reversing its
sign, $\phi \rightarrow -\phi $, produces a horizontal displacement of the
trajectory (Fig. \ref{fig:pshift_mixpure}).

Second, non-cyclic navigation arises from out-of-phase exchange modulation.
In this regime, the entanglement trajectory becomes non-periodic and is
driven away from the boundary (\ref{fig:outofp_mixpure}), enabling access to
higher entanglement values but without essential stabilization. We employ
damping to stabilize the dynamics: the exchange-time integral gradually
converges, slowing the evolution and ultimately freezing the system at a
fixed entanglement value, as illustrated in Fig. \ref{fig:dampoutofp_mixpure}%
. This final fixed entanglement is tunable through the damping strength,
phase difference, and vibrational frequencies, as indicated by the converged
ETI given in Eq. (\ref{eq:IABtinfity}). Together, cyclic and damped
non-cyclic regimes provide complementary control strategies for reversible
manipulation and robust state preparation, respectively.

In addition, the advantages of the proposed device are as follows. The
system is scalable for pairwise entanglement between qubits $A_{q}$ and $%
A_{q+1}$. For scalability, one simply relabels $A\rightarrow A_{1}$, $%
B\rightarrow A_{2}$, and $C\rightarrow C_{1}$ and then repeats the structure
to form a chain $A_{1}$-$C_{1}$-$A_{2}$-$C_{2}$-$A_{3}\cdots $-$A_{Q-1}$-$%
C_{Q-1}$-$A_{Q}$ consisting of $Q$ qubits. The eigenvalue decomposition (\ref%
{eq:VDVdag}) then becomes%
\begin{equation*}
\sum\limits_{q=1}^{Q}\sum_{k=x,y,z}\left( \sigma _{k}^{A_{q}}\eta
_{J,k}^{A_{q}}\right) S_{k}^{C_{q}}=VDV^{\dagger }\text{,}
\end{equation*}%
with $\eta _{J,k}^{A_{1}}=\gamma _{2}\eta _{J,k}^{A_{2}}=\cdots =\gamma
_{n}\eta _{J,k}^{A_{n}}$, and the same line of argument based on the ETI (%
\ref{eq:It_gen}) remains applicable. Moreover, partial correction of errors
due to dephasing is integrated into the oscillatory trajectories. The nodes (%
$J=0$) provide ideal locations for qubit information storage, while the
shapeable trajectories enable the design of quantum encryption protocols and
gate operations based on finite or repeated entanglement survival within
specific time windows, achieving on-demand entanglement. In quantum sensing
and metrology, the boundary-residing trajectory can be used for highly
precise measurements of external observables that perturb the qubit toward
or away from the entangled regime. The out-of-phase damping permits the
establishment and maintenance of strong entanglement. Accordingly, the
presented approach supports practical quantum devices that are both
efficient, by shortening entanglement generation time, and stable, through
the use of the exchange node, thereby advancing computation, secure
communication, and sensing with entanglement as a tunable resource.

\begin{acknowledgments}
One of the authors (S.-H. Chen) thanks Chang Yen Jui and Tsung-Wei Huang for valuable discussions. S-G Tan acknowledges support from the National Science and Technology Council (NSTC) of
Taiwan under Grant No. NSTC 114-2112-M-034-001. C-R Chang
acknowledges support from the NSTC under Grant No. NSTC 114-2112-M-033-005.
\end{acknowledgments}

\newpage \appendix

\section{Extended concurrence in terms of ETI for mixed states and existence
of snake trajectory}

\label{apd:E_C_ETI}

In contrast to pure states, for mixed states of the form (\ref{eq:Bell_rho0}%
), the evolution of the reduced DM $\rho \left( t\right) $ remains block
diagonal. In this appendix, we demonstrate that such a structure allows the
extended concurrence $\mathcal{C}_{E}\left( t\right) $ to be explicitly
expressed as a functional of the ETI $I(t)$, providing a direct and compact
method to track the entanglement evolution. In particular, we show that
snake trajectories can be identified analytically without invoking the ESP $%
\varepsilon $.

Consider the spin-up IS with $S^{C}=\hbar /2$ and the qubits are described
by (\ref{eq:Bell_rho0}). We set $\gamma =1$ for isotropic exchange. Using (%
\ref{eq:varrhot}) and (\ref{eq:rhoABt}) and defining the weight sums and
differences as $w_{s}=$ $w_{\alpha ^{+}}+w_{\alpha ^{-}}$ and $w_{d}=$ $%
\left\vert w_{\alpha ^{+}}-w_{\alpha ^{-}}\right\vert >0$, we introduce the
functional dependence on the ETI via 
\begin{equation}
C\left( I\right) =\cos \left( 3I\right) \text{,}
\end{equation}%
and 
\begin{equation}
k\left( I\right) =1-C\left( I\right) \in \left[ 0,2\right] \text{,}
\label{eq:k_I}
\end{equation}%
the reduced DM $\rho \left( t\right) $ then takes the form 
\begin{equation}
\begin{pmatrix}
\dfrac{9w_{s}+8w_{\beta ^{+}}k}{18} & 0 & 0 & \dfrac{w_{d}}{6}\left(
1+2e^{-3iI}\right) \\[10pt] 
0 & \dfrac{X^{+}}{18} & \dfrac{X^{-}}{18} & 0 \\[10pt] 
0 & \dfrac{X^{-}}{18} & \dfrac{X^{+}}{18} & 0 \\[10pt] 
\dfrac{w_{d}}{6}\left( 1+2e^{3iI}\right) & 0 & 0 & \dfrac{w_{s}\left(
9-4k\right) }{18}%
\end{pmatrix}%
\text{,}  \label{eq:A_rhomat}
\end{equation}%
where $X^{\pm }=X\pm 9w_{\beta ^{-}}$ and%
\begin{equation}
X=9w_{\beta ^{+}}+2k\left( w_{s}-2w_{\beta ^{+}}\right) \text{.}
\end{equation}%
Recall that $\mathcal{C}_{E}\left( t\right) $ is constructed from the
eigenvalues of (\ref{eq:sqrtR}), i.e., the square root of the eigenvalues ($%
\sqrt{\widetilde{\kappa }_{i}}=\kappa _{i}$, $i\in \{1,2,3,4\}$) of%
\begin{equation}
\sqrt{\rho \left( t\right) }\rho ^{\prime }\left( t\right) \sqrt{\rho \left(
t\right) }\text{.}  \label{eq:A_R1}
\end{equation}%
Since $\rho \left( t\right) $ and $\rho ^{\prime }\left( t\right) $ are
positive semi-definite Hermitian matrices, Eq. (\ref{eq:A_R1}) and 
\begin{equation}
R\left( t\right) =\rho \left( t\right) \rho ^{\prime }\left( t\right)
\end{equation}%
share the same eigenvalues $\widetilde{\kappa }_{i}$. Substituting (\ref%
{eq:A_rhomat}) into this expression, we obtain $R\left( t\right) $ as

\begin{widetext}
\begin{equation}
\begin{pmatrix}
\dfrac{9-4k}{162}\left( \dfrac{9}{2}(w_{s}^{2}+w_{d}^{2})+4w_{s}w_{\beta
^{+}}k\right) & 0 & 0 & \dfrac{w_{d}}{54}(1+2e^{-3iI})(9w_{s}+8w_{\beta
^{+}}k) \\[12pt] 
0 & \dfrac{X^{2}+81w_{\beta ^{-}}^{2}}{162} & \dfrac{X^{2}-81w_{\beta
^{-}}^{2}}{162} & 0 \\[12pt] 
0 & \dfrac{X^{2}-81w_{\beta ^{-}}^{2}}{162} & \dfrac{X^{2}+81w_{\beta
^{-}}^{2}}{162} & 0 \\[12pt] 
\dfrac{w_{s}w_{d}}{54}(1+2e^{3iI})(9-4k) & 0 & 0 & \dfrac{9-4k}{162}\left( 
\dfrac{9}{2}(w_{s}^{2}+w_{d}^{2})+4w_{s}w_{\beta ^{+}}k\right)%
\end{pmatrix}%
\text{.}
\end{equation}%
\end{widetext}Solving for the square roots of the eigenvalues of $R\left(
t\right) $ yields%
\begin{equation}
\kappa _{1}\left( k\right) =\frac{9w_{\beta ^{+}}+2(w_{s}-2w_{\beta ^{+}})k}{%
9}
\end{equation}%
\begin{equation}
\kappa _{2}\left( k\right) =\frac{\sqrt{9-4k}}{18}\left( \sqrt{%
w_{s}(9w_{s}+8w_{\beta ^{+}}k)}-3w_{d}\right)
\end{equation}%
\begin{equation}
\kappa _{3}\left( k\right) =\frac{\sqrt{9-4k}}{18}\left( \sqrt{%
w_{s}(9w_{s}+8w_{\beta ^{+}}k)}+3w_{d}\right)
\end{equation}%
and%
\begin{equation}
\kappa _{4}=w_{\beta ^{-}}\text{.}
\end{equation}%
Note that $\kappa _{3}\left( k\right) >\kappa _{2}\left( k\right) $ as seen
from the sign of the last term; thus, the maximum eigenvalue must be $\kappa
_{\max }\in \left\{ \kappa _{1},\kappa _{3},\kappa _{4}\right\} $.

As $k$ increases, the condition 
\begin{equation}
w_{s}>2w_{\beta ^{+}}  \label{eq:cond_s>2bm}
\end{equation}%
ensures that $\kappa _{1}\left( k\right) $ is an increasing function of $k$.
Conversely, $\kappa _{3}\left( k\right) $ decreases with $k$ because the
decay of the prefactor $\sqrt{9-4k}$ dominates the growth of $\sqrt{%
w_{s}(9w_{s}+8w_{\beta ^{+}}k)}$. The equality $\kappa _{1}\left( k\right)
=\kappa _{3}\left( k\right) $ always occurs within the interval $k\in \left[
0,2\right] $. This can be verified by analyzing the boundary values: 
\begin{equation}
\kappa _{1}\left( 0\right) =w_{\beta ^{+}}
\end{equation}%
\begin{equation}
\kappa _{3}\left( 0\right) =\frac{w_{s}+w_{d}}{2}
\end{equation}%
\begin{equation}
\kappa _{1}\left( 2\right) =\frac{4w_{s}+w_{\beta ^{+}}}{9}
\end{equation}%
\begin{equation}
\kappa _{3}\left( 2\right) =\frac{\sqrt{w_{s}\left( 9w_{s}+16w_{\beta
^{+}}\right) }+3w_{d}}{18}\text{.}
\end{equation}%
Under the condition in Eq. (\ref{eq:cond_s>2bm}), these boundaries imply $%
\kappa _{1}\left( 0\right) -\kappa _{3}\left( 0\right) <0$ and $\kappa
_{1}\left( 2\right) -\kappa _{3}\left( 2\right) >0$. For simplicity, if we
assume%
\begin{equation}
w_{\beta ^{-}}=0\text{,}  \label{eq:wbm=0}
\end{equation}%
then $\kappa _{4}=0$ and thus $\kappa _{\max }\in \left\{ \kappa _{1},\kappa
_{3}\right\} $. Under this assumption, at the crossing point where $\kappa
_{1}=\kappa _{3}$, Eq. (\ref{eq:C_E}) guarantees that $\mathcal{C}_{E}=-$ $%
\kappa _{2}<0$. Consequently, the question of whether a snake trajectory
exists reduces to determining if a positive $\mathcal{C}_{E}>0$ exists
elsewhere within $k\in \left[ 0,2\right] $. In fact, under the two
conditions (\ref{eq:cond_s>2bm}) and (\ref{eq:wbm=0}), we always find a
positive value for $\mathcal{C}_{E}$ at $k=2$, $\left. \mathcal{C}%
_{E}\right\vert _{k=2}=\kappa _{1}\left( 2\right) -\kappa _{2}\left(
2\right) -\kappa _{3}\left( 2\right) >0$, since $\kappa _{1}\left( 2\right)
>\kappa _{2}\left( 2\right) +\kappa _{3}\left( 2\right) =\sqrt{%
w_{s}(9w_{s}+16w_{\beta ^{+}})}/9$. Therefore, these two conditions---while
not optimized for maximizing the balance between the entangled and separable
residence times---are sufficient to produce a sign change in $\mathcal{C}%
_{E} $, signifying the emergence of the snake trajectory.

\begin{figure}[tbph]
\centerline{\psfig{file=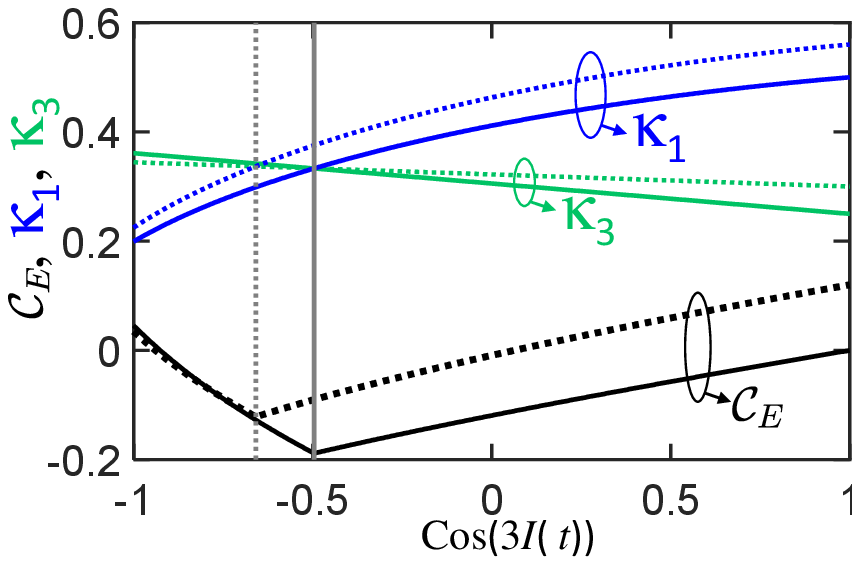,scale=0.55,angle=90,clip=true,
trim=105mm 115mm 0mm 0mm}}
\caption{Eigenvalues $\protect\kappa _{1}$ (blue) and $\protect\kappa _{3}$
(green), and extended concurrence $\mathcal{C}_{E}\left( t\right) $ (black)
plotted as functions of $\cos \left( 3I\left( t\right) \right) $. The solid
lines correspond to the weighting $\left( w_{\protect\alpha ^{+}},w_{\protect%
\alpha ^{-}},w_{\protect\beta ^{+}},w_{\protect\beta ^{-}}\right) =\left(
0.5,0.25,0.25,0\right) $, while the dotted lines use $\left( w_{\protect%
\alpha ^{+}},w_{\protect\alpha ^{-}},w_{\protect\beta ^{+}},w_{\protect\beta %
^{-}}\right) =\left( 0.56,0.14,0.3,0\right) $. Both weightings satisfy the
conditions in Eqs. (\protect\ref{eq:cond_s>2bm}) and (\protect\ref{eq:wbm=0}%
), which produce the snake trajectory. The vertical lines mark the points
where $\protect\kappa _{1}$ and $\protect\kappa _{3}$ cross.}
\label{fig:C_E_plot}
\end{figure}

The above analysis is illustrated in Fig. \ref{fig:C_E_plot}, which plots $%
\kappa _{1}$, $\kappa _{3}$, and $\mathcal{C}_{E}$ as functions of $\cos
\left( 3I\left( t\right) \right) $. Under the conditions (\ref{eq:cond_s>2bm}%
) and (\ref{eq:wbm=0}), a crossing of the eigenvalues $\kappa _{1}$ and $%
\kappa _{3}$ results in negative concurrence $\mathcal{C}_{E}<0$, whereas
positive concurrence $\mathcal{C}_{E}>0$ is recovered at $\cos \left(
3I\left( t\right) \right) =-1$ (corresponding to $k=2$), thereby realizing
the snake trajectory. By controlling the time spent in the positive and
negative $\mathcal{C}_{E}$ regions via qubit motion, a programmable
entanglement trajectory can be achieved.

\section{Analytic estimation and numerical validation of the characteristic
period}

\label{apd:Ana_est_Tstar} 
\begin{table*}[tbph]
\caption{Approximate analytic characteristic period $T^{\ast }$ as a
function of the entanglement switch parameter $\protect\varepsilon $ for 
\emph{mixed} states, obtained with \textit{time-independent} (constant) $%
J_{x}^{A/B}\left( t\right) =J_{y}^{AB}\left( t\right) =J_{z}^{AB}\left(
t\right) =J_{0}$. Equivalent weightings that yield the same analytic form of 
$T^{\ast }$ are grouped together. The expressions are valid up to order $%
O(dt^{2})$, with $dt$ the elapsed time after $t=0$. The last two columns
show the ETI for constant $J^{AB}$ compared with the ETI for sinusoidal
exchange in Table~\protect\ref{tab:Tstar}, calculated without the short-time
approximation.}
\label{tab:est_Tstar}
\centering%
\begin{tabular}{c|c|c|c}
\toprule Weightings & Approximate $T^{\ast }$ $\left( \hbar /J_{0}\right) $
& $%
\begin{array}{c}
\text{Constant $J$,} \\ 
\text{Approximate }I\left( \frac{T^{\ast }}{4}\right)%
\end{array}%
$ & $%
\begin{array}{c}
\text{Sinusoidal $J$, } \\ 
I\left( \frac{T^{\ast }}{4}\right) \text{ (Table \ref{tab:Tstar})}%
\end{array}%
$ \\ \hline
$W_{1},W_{3},W_{5}$ & $\sqrt{\dfrac{16\varepsilon }{1+\varepsilon }}$ & 0.157
& 0.100 \\ \hline
$W_{2},W_{4}$ & $2\sqrt{2}$ & 0.655 & 0.417 \\ \hline
$W_{7},W_{8}$ & $\sqrt{\dfrac{32\varepsilon }{1+3\varepsilon }}$ & 0.220 & 
0.140 \\ \hline
$W_{9}$ & $\sqrt{\dfrac{32\varepsilon }{3+5\varepsilon }}$ & 0.129 & 0.082
\\ \hline
$W_{10}$ & $\left[ -\dfrac{1024\varepsilon (1+\varepsilon )}{(1-\varepsilon
)^{2}}\right] ^{1/4}$ & 0.721 & 0.459 \\ \hline
$W_{11},W_{12}$ & $\sqrt{\dfrac{24\varepsilon }{(1+2\varepsilon )}}$ & 0.191
& 0.122 \\ \hline
$W_{13}$ & $\sqrt{\dfrac{12\varepsilon }{(1+2\varepsilon )}}$ & 0.136 & 0.087
\\ \hline
$W_{14}$ & $\left[ -\dfrac{(384+768\varepsilon )\varepsilon }{(1-\varepsilon
)^{2}}\right] ^{1/4}$ & 0.575 & 0.366 \\ 
\bottomrule
\end{tabular}%
\end{table*}

For mixed states, the characteristic period $T^{\ast }$, defined by the
separability-boundary criterion $\mathcal{C}_{E}(T^{\ast }/4)=0$, can be
estimated analytically under the short-time approximation. Based on the
results in Table 2 of Physica Scripta 100, 065114 (2025), which assumes a
constant exchange interaction $J^{A/B}\left( t\right) =J^{A/B}$, we compute
the analytic expressions for $T^{\ast }$ valid up to order $O\left(
dt^{2}\right) $. These approximate values are listed in Table \ref%
{tab:est_Tstar}. To validate this approximation for our
alternating-RKKY-exchange scheme, we compare the ETI calculated using these
constant-exchange analytic periods against the full numerical results for
sinusoidal exchange (as adopted in the main text). As shown in the last two
columns of Table \ref{tab:est_Tstar}, the constant-exchange ETI is in good
agreement with the exact numerical values, confirming that the short-time
analytic model provides a reliable estimate for the near-boundary dynamics
studied herein.


\bibliographystyle{naturemag}
\bibliography{2025shc}

@ARTICLE{Kasuya1956,
  author = {T. Kasuya},
  title = {A theory of metallic ferro- and antiferromagnetism},
  journal = {Progr. Theoret. Phys.},
  year = {1956},
  volume = {16},
  pages = {45}
}

@ARTICLE{Ruderman1954,
  author = {M. A. Ruderman and C. Kittel},
  title = {Indirect Exchange Coupling of Nuclear Magnetic Moments by Conduction
	Electrons},
  journal = {Phys. Rev.},
  year = {1954},
  volume = {96},
  pages = {99}
}

@ARTICLE{Yosida1957,
  author = {Kei Yosida},
  title = {Magnetic Properties of Cu-Mn Alloys},
  journal = {Phys. Rev.},
  year = {1957},
  volume = {106},
  pages = {893}
}

@Article{Khalili2018,
  author    = {Khalili, F. Ya. and Polzik, E. S.},
  journal   = {Phys. Rev. Lett.},
  title     = {Overcoming the Standard Quantum Limit in Gravitational Wave Detectors Using Spin Systems with a Negative Effective Mass},
  year      = {2018},
  month     = {Jul},
  pages     = {031101},
  volume    = {121},
  doi       = {10.1103/PhysRevLett.121.031101},
  issue     = {3},
  numpages  = {5},
  publisher = {American Physical Society},
  url       = {https://link.aps.org/doi/10.1103/PhysRevLett.121.031101},
}

@Article{Zeuthen2019,
  author    = {Zeuthen, Emil and Polzik, Eugene S and Khalili, Farid Ya},
  journal   = {Physical Review D},
  title     = {Gravitational wave detection beyond the standard quantum limit using a negative-mass spin system and virtual rigidity},
  year      = {2019},
  number    = {6},
  pages     = {062004},
  volume    = {100},
  publisher = {APS},
}

@Article{Horodecki2009,
  author    = {Horodecki, Ryszard and Horodecki, Pawe\l{} and Horodecki, Micha\l{} and Horodecki, Karol},
  journal   = {Rev. Mod. Phys.},
  title     = {Quantum entanglement},
  year      = {2009},
  month     = {Jun},
  pages     = {865--942},
  volume    = {81},
  doi       = {10.1103/RevModPhys.81.865},
  issue     = {2},
  numpages  = {0},
  publisher = {American Physical Society},
  url       = {https://link.aps.org/doi/10.1103/RevModPhys.81.865},
}

@Article{Shor2000,
  author    = {Shor, Peter W and Preskill, John},
  journal   = {Physical review letters},
  title     = {Simple proof of security of the BB84 quantum key distribution protocol},
  year      = {2000},
  number    = {2},
  pages     = {441},
  volume    = {85},
  publisher = {APS},
}

@Article{Gisin2002,
  author    = {Gisin, Nicolas and Ribordy, Gr{\'e}goire and Tittel, Wolfgang and Zbinden, Hugo},
  journal   = {Reviews of modern physics},
  title     = {Quantum cryptography},
  year      = {2002},
  number    = {1},
  pages     = {145},
  volume    = {74},
  publisher = {APS},
}

@Article{Bennett1998,
  author    = {Bennett, Charles H},
  journal   = {Physica Scripta},
  title     = {Quantum information},
  year      = {1998},
  number    = {T76},
  pages     = {210},
  volume    = {1998},
  publisher = {IOP Publishing},
}

@Article{Steane1998,
  author    = {Steane, Andrew},
  journal   = {Reports on Progress in Physics},
  title     = {Quantum computing},
  year      = {1998},
  number    = {2},
  pages     = {117},
  volume    = {61},
  publisher = {IOP Publishing},
}

@Article{DiVincenzo1995,
  author    = {DiVincenzo, David P},
  journal   = {Science},
  title     = {Quantum computation},
  year      = {1995},
  number    = {5234},
  pages     = {255--261},
  volume    = {270},
  publisher = {American Association for the Advancement of Science},
}

@Article{DiVincenzo2000,
  author    = {DiVincenzo, David P},
  journal   = {Fortschritte der Physik: Progress of Physics},
  title     = {The physical implementation of quantum computation},
  year      = {2000},
  number    = {9-11},
  pages     = {771--783},
  volume    = {48},
  publisher = {Wiley Online Library},
}

@Article{Ladd2010,
  author    = {Ladd, Thaddeus D and Jelezko, Fedor and Laflamme, Raymond and Nakamura, Yasunobu and Monroe, Christopher and O Brien, Jeremy Lloyd},
  journal   = {nature},
  title     = {Quantum computers},
  year      = {2010},
  number    = {7285},
  pages     = {45--53},
  volume    = {464},
  publisher = {Nature Publishing Group UK London},
}

@Article{Madsen2022,
  author    = {Madsen, Lars S and Laudenbach, Fabian and Askarani, Mohsen Falamarzi and Rortais, Fabien and Vincent, Trevor and Bulmer, Jacob FF and Miatto, Filippo M and Neuhaus, Leonhard and Helt, Lukas G and Collins, Matthew J and others},
  journal   = {Nature},
  title     = {Quantum computational advantage with a programmable photonic processor},
  year      = {2022},
  number    = {7912},
  pages     = {75--81},
  volume    = {606},
  publisher = {Nature Publishing Group UK London},
}

@Article{Hill1997,
  author    = {Hill, Sam A. and Wootters, William K.},
  journal   = {Phys. Rev. Lett.},
  title     = {Entanglement of a Pair of Quantum Bits},
  year      = {1997},
  month     = {Jun},
  pages     = {5022--5025},
  volume    = {78},
  doi       = {10.1103/PhysRevLett.78.5022},
  issue     = {26},
  numpages  = {0},
  publisher = {American Physical Society},
  url       = {https://link.aps.org/doi/10.1103/PhysRevLett.78.5022},
}

@Article{Rungta2001,
  author    = {Rungta, Pranaw and Bu\ifmmode \check{z}\else \v{z}\fi{}ek, V. and Caves, Carlton M. and Hillery, M. and Milburn, G. J.},
  journal   = {Phys. Rev. A},
  title     = {Universal state inversion and concurrence in arbitrary dimensions},
  year      = {2001},
  month     = {Sep},
  pages     = {042315},
  volume    = {64},
  doi       = {10.1103/PhysRevA.64.042315},
  issue     = {4},
  numpages  = {13},
  publisher = {American Physical Society},
  url       = {https://link.aps.org/doi/10.1103/PhysRevA.64.042315},
}

@Article{Wootters1998,
  author    = {Wootters, William K.},
  journal   = {Phys. Rev. Lett.},
  title     = {Entanglement of formation of an arbitrary state of two qubits},
  year      = {1998},
  month     = {Mar},
  pages     = {2245--2248},
  volume    = {80},
  doi       = {10.1103/PhysRevLett.80.2245},
  issue     = {10},
  numpages  = {0},
  publisher = {American Physical Society},
  url       = {https://link.aps.org/doi/10.1103/PhysRevLett.80.2245},
}

@Article{Mondal2021,
  author    = {Mondal, Priyanka and Suresh, Abhin and Nikoli\ifmmode \acute{c}\else \'{c}\fi{}, Branislav K.},
  journal   = {Phys. Rev. B},
  title     = {When can localized spins interacting with conduction electrons in ferro- or antiferromagnets be described classically via the Landau-Lifshitz equation: Transition from quantum many-body entangled to quantum-classical nonequilibrium states},
  year      = {2021},
  month     = {Dec},
  pages     = {214401},
  volume    = {104},
  doi       = {10.1103/PhysRevB.104.214401},
  issue     = {21},
  numpages  = {10},
  publisher = {American Physical Society},
  url       = {https://link.aps.org/doi/10.1103/PhysRevB.104.214401},
}

@Article{Wang2018,
  author    = {Wang, F. and Hou, P.-Y. and Huang, Y.-Y. and Zhang, W.-G. and Ouyang, X.-L. and Wang, X. and Huang, X.-Z. and Zhang, H.-L. and He, L. and Chang, X.-Y. and Duan, L.-M.},
  journal   = {Phys. Rev. B},
  title     = {Observation of entanglement sudden death and rebirth by controlling a solid-state spin bath},
  year      = {2018},
  month     = {Aug},
  pages     = {064306},
  volume    = {98},
  doi       = {10.1103/PhysRevB.98.064306},
  issue     = {6},
  numpages  = {7},
  publisher = {American Physical Society},
  url       = {https://link.aps.org/doi/10.1103/PhysRevB.98.064306},
}

@Article{Yu2004a,
  author    = {Yu, Ting and Eberly, J. H.},
  journal   = {Phys. Rev. Lett.},
  title     = {Finite-Time Disentanglement Via Spontaneous Emission},
  year      = {2004},
  month     = {Sep},
  pages     = {140404},
  volume    = {93},
  doi       = {10.1103/PhysRevLett.93.140404},
  issue     = {14},
  numpages  = {4},
  publisher = {American Physical Society},
  url       = {https://link.aps.org/doi/10.1103/PhysRevLett.93.140404},
}

@Article{Almeida2007,
  author    = {Almeida, Marcelo P and de Melo, Fernando and Hor-Meyll, Malena and Salles, Alejo and Walborn, SP and Ribeiro, PH Souto and Davidovich, Luiz},
  journal   = {science},
  title     = {Environment-induced sudden death of entanglement},
  year      = {2007},
  number    = {5824},
  pages     = {579--582},
  volume    = {316},
  publisher = {American Association for the Advancement of Science},
}

@Article{Yu2009,
  author    = {Ting Yu and J. H. Eberly},
  journal   = {Science},
  title     = {Sudden death of entanglement},
  year      = {2009},
  number    = {5914},
  pages     = {598--601},
  volume    = {323},
  publisher = {American Association for the Advancement of Science},
}

@Article{Zhang2024,
  author    = {Zhang, Zheshen and You, Chenglong and Maga{\~n}a-Loaiza, Omar S and Fickler, Robert and Le{\'o}n-Montiel, Roberto de J and Torres, Juan P and Humble, Travis S and Liu, Shuai and Xia, Yi and Zhuang, Quntao},
  journal   = {Advances in Optics and Photonics},
  title     = {Entanglement-based quantum information technology: a tutorial},
  year      = {2024},
  number    = {1},
  pages     = {60--162},
  volume    = {16},
  publisher = {Optica Publishing Group},
}

@Article{Chen2024,
  author    = {Chen, Son-Hsien},
  journal   = {Phys. Rev. B},
  title     = {Origin and early growth of entanglement by $sd$ exchange with gate voltage controllable outcome},
  year      = {2024},
  month     = {Jan},
  pages     = {045308},
  volume    = {109},
  doi       = {10.1103/PhysRevB.109.045308},
  issue     = {4},
  numpages  = {14},
  publisher = {American Physical Society},
  url       = {https://link.aps.org/doi/10.1103/PhysRevB.109.045308},
}

@Article{Bennett2000,
  author    = {Bennett, Charles H and DiVincenzo, David P},
  journal   = {nature},
  title     = {Quantum information and computation},
  year      = {2000},
  number    = {6775},
  pages     = {247--255},
  volume    = {404},
  publisher = {Nature Publishing Group UK London},
}

@Book{Nielsen2010,
  author    = {M. A. Nielsen and I. L. Chuang},
  publisher = {Cambridge university press},
  title     = {Quantum computation and quantum information},
  year      = {2010},
}

@Article{Yu2006a,
  author    = {Yu, Ting and Eberly, JH},
  journal   = {Optics Communications},
  title     = {Sudden death of entanglement: classical noise effects},
  year      = {2006},
  number    = {2},
  pages     = {393--397},
  volume    = {264},
  publisher = {Elsevier},
}

@Article{Yu2006b,
  author    = {Yu, Ting and Eberly, JH},
  journal   = {Physical review letters},
  title     = {Quantum open system theory: bipartite aspects},
  year      = {2006},
  number    = {14},
  pages     = {140403},
  volume    = {97},
  publisher = {APS},
}

@Article{Ann2007,
  author    = {Ann, Kevin and Jaeger, Gregg},
  journal   = {Phys. Rev. A},
  title     = {Local-dephasing-induced entanglement sudden death in two-component finite-dimensional systems},
  year      = {2007},
  month     = {Oct},
  pages     = {044101},
  volume    = {76},
  doi       = {10.1103/PhysRevA.76.044101},
  issue     = {4},
  numpages  = {4},
  publisher = {American Physical Society},
  url       = {https://link.aps.org/doi/10.1103/PhysRevA.76.044101},
}

@Article{ifmmodeZelseZfiyczkowski2001,
  author    = {\ifmmode \dot{Z}\else \.{Z}\fi{}yczkowski, Karol and Horodecki, Pawe\l{} and Horodecki, Micha\l{} and Horodecki, Ryszard},
  journal   = {Phys. Rev. A},
  title     = {Dynamics of quantum entanglement},
  year      = {2001},
  month     = {Dec},
  pages     = {012101},
  volume    = {65},
  doi       = {10.1103/PhysRevA.65.012101},
  issue     = {1},
  numpages  = {9},
  publisher = {American Physical Society},
  url       = {https://link.aps.org/doi/10.1103/PhysRevA.65.012101},
}

@Article{GarciaGaitan2024,
  author    = {Garcia-Gaitan, Federico and Nikoli\ifmmode \acute{c}\else \'{c}\fi{}, Branislav K.},
  journal   = {Phys. Rev. B},
  title     = {Fate of entanglement in magnetism under {L}indbladian or non-{M}arkovian dynamics and conditions for their transition to {L}andau-{L}ifshitz-{G}ilbert classical dynamics},
  year      = {2024},
  month     = {May},
  pages     = {L180408},
  volume    = {109},
  doi       = {10.1103/PhysRevB.109.L180408},
  issue     = {18},
  numpages  = {8},
  publisher = {American Physical Society},
  url       = {https://link.aps.org/doi/10.1103/PhysRevB.109.L180408},
}

@Article{Hutton2004,
  author    = {Hutton, A. and Bose, S.},
  journal   = {Phys. Rev. A},
  title     = {Mediated entanglement and correlations in a star network of interacting spins},
  year      = {2004},
  month     = {Apr},
  pages     = {042312},
  volume    = {69},
  doi       = {10.1103/PhysRevA.69.042312},
  issue     = {4},
  numpages  = {7},
  publisher = {American Physical Society},
  url       = {https://link.aps.org/doi/10.1103/PhysRevA.69.042312},
}

@Article{Yuan2007,
  author    = {Yuan, Xiao-Zhong and Goan, Hsi-Sheng and Zhu, Ka-Di},
  journal   = {Phys. Rev. B},
  title     = {Non-{M}arkovian reduced dynamics and entanglement evolution of two coupled spins in a quantum spin environment},
  year      = {2007},
  month     = {Jan},
  pages     = {045331},
  volume    = {75},
  doi       = {10.1103/PhysRevB.75.045331},
  issue     = {4},
  numpages  = {8},
  publisher = {American Physical Society},
  url       = {https://link.aps.org/doi/10.1103/PhysRevB.75.045331},
}

@Article{Sharma2021,
  author    = {Sharma, Amritesh and Tulapurkar, Ashwin A},
  journal   = {Physical Review A},
  title     = {Transmission-based tomography for spin qubits},
  year      = {2021},
  number    = {5},
  pages     = {052430},
  volume    = {103},
  publisher = {APS},
}

@Article{CostaJr2001,
  author    = {Costa Jr, AT and Bose, S},
  journal   = {Physical review letters},
  title     = {Impurity scattering induced entanglement of ballistic electrons},
  year      = {2001},
  number    = {27},
  pages     = {277901},
  volume    = {87},
  publisher = {APS},
}

@Article{Mortezapour2017,
  author    = {Mortezapour, Ali and Borji, Mahdi Ahmadi and Franco, Rosario Lo},
  journal   = {Laser Physics Letters},
  title     = {Protecting entanglement by adjusting the velocities of moving qubits inside non-Markovian environments},
  year      = {2017},
  number    = {5},
  pages     = {055201},
  volume    = {14},
  publisher = {IOP Publishing},
}

@Article{Trenyi2024,
  author    = {Tr{\'e}nyi, R{\'o}bert and Luk{\'a}cs, {\'A}rp{\'a}d and Horodecki, Pawe{\l} and Horodecki, Ryszard and V{\'e}rtesi, Tam{\'a}s and T{\'o}th, G{\'e}za},
  journal   = {New Journal of Physics},
  title     = {Activation of metrologically useful genuine multipartite entanglement},
  year      = {2024},
  number    = {2},
  pages     = {023034},
  volume    = {26},
  publisher = {IOP Publishing},
}

@Article{Ekert1991,
  author    = {Ekert, Artur K.},
  journal   = {Phys. Rev. Lett.},
  title     = {Quantum cryptography based on Bell's theorem},
  year      = {1991},
  month     = {Aug},
  pages     = {661--663},
  volume    = {67},
  doi       = {10.1103/PhysRevLett.67.661},
  issue     = {6},
  numpages  = {0},
  publisher = {American Physical Society},
  url       = {https://link.aps.org/doi/10.1103/PhysRevLett.67.661},
}

@Article{Wang2025,
  author  = {Wang, Tian-Le and Wang, Peng and Zhao, Ze-An and Zhang, Sheng and Zhao, Ren-Ze and Yang, Xiao-Yan and Zhang, Hai-Feng and Li, Zhi-Fei and Wu, Yuan and Guo, Liang-Liang and others},
  journal = {arXiv preprint arXiv:2506.06669},
  title   = {Remote entanglement generation via enhanced quantum state transfer},
  year    = {2025},
}

@Article{RieraSabat2023,
  author    = {Riera-S{\`a}bat, Ferran and Sekatski, Pavel and D{\"u}r, Wolfgang},
  journal   = {Quantum},
  title     = {Remotely controlled entanglement generation},
  year      = {2023},
  pages     = {904},
  volume    = {7},
  publisher = {Verein zur F{\"o}rderung des Open Access Publizierens in den Quantenwissenschaften},
}

@Article{Preskill2018,
  author    = {Preskill, John},
  journal   = {Quantum},
  title     = {Quantum computing in the NISQ era and beyond},
  year      = {2018},
  pages     = {79},
  volume    = {2},
  publisher = {Verein zur F{\"o}rderung des Open Access Publizierens in den Quantenwissenschaften},
}

@Article{Cirac1995,
  author    = {Cirac, J. I. and Zoller, P.},
  journal   = {Phys. Rev. Lett.},
  title     = {Quantum Computations with Cold Trapped Ions},
  year      = {1995},
  month     = {May},
  pages     = {4091--4094},
  volume    = {74},
  doi       = {10.1103/PhysRevLett.74.4091},
  issue     = {20},
  numpages  = {0},
  publisher = {American Physical Society},
  url       = {https://link.aps.org/doi/10.1103/PhysRevLett.74.4091},
}

@Article{Monroe2014,
  author    = {Monroe, Christopher and Raussendorf, Robert and Ruthven, Alex and Brown, Kenneth R and Maunz, Peter and Duan, L-M and Kim, Jungsang},
  journal   = {Physical Review A},
  title     = {Large-scale modular quantum-computer architecture with atomic memory and photonic interconnects},
  year      = {2014},
  number    = {2},
  pages     = {022317},
  volume    = {89},
  publisher = {APS},
}

@Article{Tyryshkin2012,
  author    = {Tyryshkin, Alexei M and Tojo, Shinichi and Morton, John JL and Riemann, Helge and Abrosimov, Nikolai V and Becker, Peter and Pohl, Hans-Joachim and Schenkel, Thomas and Thewalt, Michael LW and Itoh, Kohei M and others},
  journal   = {Nature materials},
  title     = {Electron spin coherence exceeding seconds in high-purity silicon},
  year      = {2012},
  number    = {2},
  pages     = {143--147},
  volume    = {11},
  publisher = {Nature Publishing Group UK London},
}

@Article{Pla2013,
  author    = {Pla, Jarryd J and Tan, Kuan Y and Dehollain, Juan P and Lim, Wee H and Morton, John JL and Zwanenburg, Floris A and Jamieson, David N and Dzurak, Andrew S and Morello, Andrea},
  journal   = {Nature},
  title     = {High-fidelity readout and control of a nuclear spin qubit in silicon},
  year      = {2013},
  number    = {7445},
  pages     = {334--338},
  volume    = {496},
  publisher = {Nature Publishing Group UK London},
}

@Article{Loss1998,
  author    = {Loss, Daniel and DiVincenzo, David P.},
  journal   = {Phys. Rev. A},
  title     = {Quantum computation with quantum dots},
  year      = {1998},
  month     = {Jan},
  pages     = {120--126},
  volume    = {57},
  doi       = {10.1103/PhysRevA.57.120},
  issue     = {1},
  numpages  = {0},
  publisher = {American Physical Society},
  url       = {https://link.aps.org/doi/10.1103/PhysRevA.57.120},
}

@Article{Nowack2007,
  author    = {Nowack, Katja C and Koppens, FHL and Nazarov, Yu V and Vandersypen, LMK},
  journal   = {Science},
  title     = {Coherent control of a single electron spin with electric fields},
  year      = {2007},
  number    = {5855},
  pages     = {1430--1433},
  volume    = {318},
  publisher = {American Association for the Advancement of Science},
}

@Article{Petta2005,
  author  = {J. R. Petta and A. C. Johnson and J. M. Taylor and E. A. Laird and A. Yacoby and M. D. Lukin and C. M. Marcus and M. P. Hanson and A. C. Gossard},
  journal = {Science},
  title   = {Coherent manipulation of coupled electron spins in semiconductor quantum dots},
  year    = {2005},
  note    = {Epub 2005 Sep 1},
  number  = {5744},
  pages   = {2180--2184},
  volume  = {309},
  doi     = {10.1126/science.1116955},
}

@Article{Bluhm2011,
  author    = {Bluhm, Hendrik and Foletti, Sandra and Neder, Izhar and Rudner, Mark and Mahalu, Diana and Umansky, Vladimir and Yacoby, Amir},
  journal   = {Nature Physics},
  title     = {Dephasing time of GaAs electron-spin qubits coupled to a nuclear bath exceeding 200 $\mu$s},
  year      = {2011},
  number    = {2},
  pages     = {109--113},
  volume    = {7},
  publisher = {Nature Publishing Group UK London},
}

@Article{Arute2019,
  author    = {Arute, Frank and Arya, Kunal and Babbush, Ryan and Bacon, Dave and Bardin, Joseph C and Barends, Rami and Biswas, Rupak and Boixo, Sergio and Brandao, Fernando GSL and Buell, David A and others},
  journal   = {Nature},
  title     = {Quantum supremacy using a programmable superconducting processor},
  year      = {2019},
  number    = {7779},
  pages     = {505--510},
  volume    = {574},
  publisher = {Nature Publishing Group UK London},
}

@Article{Zhong2020,
  author    = {Zhong, Han-Sen and Wang, Hui and Deng, Yu-Hao and Chen, Ming-Cheng and Peng, Li-Chao and Luo, Yi-Han and Qin, Jian and Wu, Dian and Ding, Xing and Hu, Yi and others},
  journal   = {Science},
  title     = {Quantum computational advantage using photons},
  year      = {2020},
  number    = {6523},
  pages     = {1460--1463},
  volume    = {370},
  publisher = {American Association for the Advancement of Science},
}

@Article{Humphreys2018,
  author    = {Humphreys, Peter C and Kalb, Norbert and Morits, Jaco PJ and Schouten, Raymond N and Vermeulen, Raymond FL and Twitchen, Daniel J and Markham, Matthew and Hanson, Ronald},
  journal   = {Nature},
  title     = {Deterministic delivery of remote entanglement on a quantum network},
  year      = {2018},
  number    = {7709},
  pages     = {268--273},
  volume    = {558},
  publisher = {Nature Publishing Group UK London},
}

@Article{Watson2018,
  author    = {Watson, Thomas F and Philips, SGJ and Kawakami, Erika and Ward, Daniel R and Scarlino, Pasquale and Veldhorst, Menno and Savage, Donald E and Lagally, MG and Friesen, Mark and Coppersmith, Susan N and others},
  journal   = {nature},
  title     = {A programmable two-qubit quantum processor in silicon},
  year      = {2018},
  number    = {7698},
  pages     = {633--637},
  volume    = {555},
  publisher = {Nature Publishing Group UK London},
}

@Article{Zajac2018,
  author    = {Zajac, David M and Sigillito, Anthony J and Russ, Maximilian and Borjans, Felix and Taylor, Jacob M and Burkard, Guido and Petta, Jason R},
  journal   = {Science},
  title     = {Resonantly driven CNOT gate for electron spins},
  year      = {2018},
  number    = {6374},
  pages     = {439--442},
  volume    = {359},
  publisher = {American Association for the Advancement of Science},
}

@Article{Lin2025,
  author  = {Lin, Li-Che and Tan, Seng Ghee and Chang, Ching-Ray and Sun, Shih-Jye and Chen, Son-Hsien},
  journal = {New Journal of Physics},
  title   = {Entanglement induced by Heisenberg exchange between an electron in a nested quantum dot and a qubit with relative motion},
  year    = {2025},
}

@Article{Chen2025,
  author    = {Chen, Son-Hsien and Tan, Seng Ghee and Huang, Che-Chun},
  journal   = {Physica Scripta},
  title     = {General recipe for immediate entanglement death and birth via Bell states: environmental Heisenberg exchange with transition as an example},
  year      = {2025},
  number    = {6},
  pages     = {065114},
  volume    = {100},
  publisher = {IOP Publishing},
}

@Article{Chen2009a,
  author    = {Chen, Son-Hsien and Maekawa, Sadamichi and Liu, Ming-Hao and Chang, Ching-Ray},
  journal   = {Journal of Physics D: Applied Physics},
  title     = {Mirror symmetry and exchange of magnetic impurities mediated by electrons of Rashba spin--orbit interaction in a four-terminal Landauer setup},
  year      = {2009},
  number    = {1},
  pages     = {015003},
  volume    = {43},
  publisher = {IOP Publishing},
}

@Article{Stocker2024,
  author    = {Stocker, Lidia and Zilberberg, Oded},
  journal   = {Physical Review Research},
  title     = {Coherent exchange-coupled nonlocal Kondo impurities},
  year      = {2024},
  number    = {2},
  pages     = {L022058},
  volume    = {6},
  publisher = {APS},
}

@Article{Kettemann2024,
  author  = {Kettemann, Stefan},
  journal = {arXiv preprint arXiv:2408.03112},
  title   = {Competition between Kondo Effect and RKKY Coupling},
  year    = {2024},
}

@Article{Mousavi2021,
  author    = {Mousavi, Fatemeh Mazhari and Farghadan, Rouhollah},
  journal   = {Journal of Physics and Chemistry of Solids},
  title     = {Electrical control of Ruderman--Kittel--Kasuya--Yosida exchange interaction in zigzag edge MoS2 nanoflakes},
  year      = {2021},
  pages     = {110242},
  volume    = {158},
  publisher = {Elsevier},
}

@Article{Utsumi2004,
  author    = {Utsumi, Yasuhiro and Martinek, Jan and Bruno, Patrick and Imamura, Hiroshi},
  journal   = {Physical Review B},
  title     = {Indirect exchange interaction between two quantum dots in an Aharonov-Bohm ring},
  year      = {2004},
  number    = {15},
  pages     = {155320},
  volume    = {69},
  publisher = {APS},
}

@Article{Cho2006,
  author    = {Cho, Sam Young and McKenzie, Ross H.},
  journal   = {Phys. Rev. A},
  title     = {Quantum entanglement in the two-impurity Kondo model},
  year      = {2006},
  month     = {Jan},
  pages     = {012109},
  volume    = {73},
  doi       = {10.1103/PhysRevA.73.012109},
  issue     = {1},
  numpages  = {7},
  publisher = {American Physical Society},
  url       = {https://link.aps.org/doi/10.1103/PhysRevA.73.012109},
}

@Article{Wang2022,
  author    = {Wang, Jia-Ning and Zhou, Wang-Huai and Yan, Yu-Xiong and Li, Wei and Nan, Nan and Zhang, Jun and Ma, Ya-Nan and Wang, Peng-Chao and Ma, Xiang-Rui and Luo, Shi-Jun and Xiong, Yong-Chen},
  journal   = {Phys. Rev. B},
  title     = {Unified formulations for RKKY interaction, side Kondo behavior, and Fano antiresonance in a hybrid tripartite quantum dot device with filtered density of states},
  year      = {2022},
  month     = {Jul},
  pages     = {035428},
  volume    = {106},
  doi       = {10.1103/PhysRevB.106.035428},
  issue     = {3},
  numpages  = {15},
  publisher = {American Physical Society},
  url       = {https://link.aps.org/doi/10.1103/PhysRevB.106.035428},
}

@Article{Vonhoff2022,
  author    = {Vonhoff, Frederik and Fischer, Andreas and Deltenre, Kira and Anders, Frithjof B.},
  journal   = {Phys. Rev. Lett.},
  title     = {Microscopic Origin of the Effective Spin-Spin Interaction in a Semiconductor Quantum Dot Ensemble},
  year      = {2022},
  month     = {Oct},
  pages     = {167701},
  volume    = {129},
  doi       = {10.1103/PhysRevLett.129.167701},
  issue     = {16},
  numpages  = {6},
  publisher = {American Physical Society},
  url       = {https://link.aps.org/doi/10.1103/PhysRevLett.129.167701},
}

@Article{Vavilov2005,
  author    = {Vavilov, Maxim G. and Glazman, Leonid I.},
  journal   = {Phys. Rev. Lett.},
  title     = {Transport Spectroscopy of Kondo Quantum Dots Coupled by RKKY Interaction},
  year      = {2005},
  month     = {Mar},
  pages     = {086805},
  volume    = {94},
  doi       = {10.1103/PhysRevLett.94.086805},
  issue     = {8},
  numpages  = {4},
  publisher = {American Physical Society},
  url       = {https://link.aps.org/doi/10.1103/PhysRevLett.94.086805},
}

@Article{Yang2016,
  author    = {Yang, Guang and Hsu, Chen-Hsuan and Stano, Peter and Klinovaja, Jelena and Loss, Daniel},
  journal   = {Phys. Rev. B},
  title     = {Long-distance entanglement of spin qubits via quantum Hall edge states},
  year      = {2016},
  month     = {Feb},
  pages     = {075301},
  volume    = {93},
  doi       = {10.1103/PhysRevB.93.075301},
  issue     = {7},
  numpages  = {15},
  publisher = {American Physical Society},
  url       = {https://link.aps.org/doi/10.1103/PhysRevB.93.075301},
}

@Article{Allerdt2015,
  author    = {Allerdt, Andrew and B\"usser, C. A. and Martins, G. B. and Feiguin, A. E.},
  journal   = {Phys. Rev. B},
  title     = {Kondo versus indirect exchange: Role of lattice and actual range of RKKY interactions in real materials},
  year      = {2015},
  month     = {Feb},
  pages     = {085101},
  volume    = {91},
  doi       = {10.1103/PhysRevB.91.085101},
  issue     = {8},
  numpages  = {7},
  publisher = {American Physical Society},
  url       = {https://link.aps.org/doi/10.1103/PhysRevB.91.085101},
}

@Article{Leon2019,
  author    = {Leon, Alejandro O. and d'Albuquerque e Castro, Jose and Retamal, Juan C. and Cahaya, Adam B. and Altbir, Dora},
  journal   = {Phys. Rev. B},
  title     = {Manipulation of the RKKY exchange by voltages},
  year      = {2019},
  month     = {Jul},
  pages     = {014403},
  volume    = {100},
  doi       = {10.1103/PhysRevB.100.014403},
  issue     = {1},
  numpages  = {6},
  publisher = {American Physical Society},
  url       = {https://link.aps.org/doi/10.1103/PhysRevB.100.014403},
}

@Article{Tran2024,
  author    = {Tran, Bao Xuan and Ha, Jae-Hyun and Choi, Won-Chang and Yoon, Seongsoo and Kim, Tae-Hwan and Hong, Jung-Il},
  journal   = {Applied Physics Letters},
  title     = {Field-free control and switching of perpendicular magnetization by voltage induced manipulation of RKKY interaction},
  year      = {2024},
  number    = {11},
  volume    = {124},
  publisher = {AIP Publishing},
}

@Article{Elman2017,
  author    = {Elman, Samuel J. and Bartlett, Stephen D. and Doherty, Andrew C.},
  journal   = {Phys. Rev. B},
  title     = {Long-range entanglement for spin qubits via quantum Hall edge modes},
  year      = {2017},
  month     = {Sep},
  pages     = {115407},
  volume    = {96},
  doi       = {10.1103/PhysRevB.96.115407},
  issue     = {11},
  numpages  = {12},
  publisher = {American Physical Society},
  url       = {https://link.aps.org/doi/10.1103/PhysRevB.96.115407},
}

@Article{Petersson2012,
  author    = {Petersson, Karl D and McFaul, Louis W and Schroer, Michael D and Jung, Minkyung and Taylor, Jacob M and Houck, Andrew A and Petta, Jason R},
  journal   = {Nature},
  title     = {Circuit quantum electrodynamics with a spin qubit},
  year      = {2012},
  number    = {7420},
  pages     = {380--383},
  volume    = {490},
  publisher = {Nature Publishing Group UK London},
}

@Article{Huan2015,
  author    = {Huan, Tiantian and Zhou, Rigui and Ian, Hou},
  journal   = {Phys. Rev. A},
  title     = {Dynamic entanglement transfer in a double-cavity optomechanical system},
  year      = {2015},
  month     = {Aug},
  pages     = {022301},
  volume    = {92},
  doi       = {10.1103/PhysRevA.92.022301},
  issue     = {2},
  numpages  = {7},
  publisher = {American Physical Society},
  url       = {https://link.aps.org/doi/10.1103/PhysRevA.92.022301},
}

@Article{Obada2010,
  author    = {Obada, AS F and Hessian, HA and Hashem, M},
  journal   = {Physica Scripta},
  title     = {Quantum entanglement in a system of two moving atoms interacting with a single mode field},
  year      = {2010},
  number    = {5},
  pages     = {055303},
  volume    = {81},
  publisher = {IOP Publishing},
}

@Article{Pandit2018,
  author    = {Pandit, Mahasweta and Das, Sreetama and Roy, Sudipto Singha and Dhar, Himadri Shekhar and Sen, Ujjwal},
  journal   = {Journal of Physics B: Atomic, Molecular and Optical Physics},
  title     = {Effects of cavity-cavity interaction on the entanglement dynamics of a generalized double Jaynes--Cummings model},
  year      = {2018},
  number    = {4},
  pages     = {045501},
  volume    = {51},
  publisher = {IOP Publishing},
}

@Article{Stocker2022,
  author    = {Stocker, Lidia and Sack, Stefan H. and Ferguson, Michael S. and Zilberberg, Oded},
  journal   = {Phys. Rev. Res.},
  title     = {Entanglement-based observables for quantum impurities},
  year      = {2022},
  month     = {Dec},
  pages     = {043177},
  volume    = {4},
  doi       = {10.1103/PhysRevResearch.4.043177},
  issue     = {4},
  numpages  = {9},
  publisher = {American Physical Society},
  url       = {https://link.aps.org/doi/10.1103/PhysRevResearch.4.043177},
}

@Article{Bazhanov2018,
  author    = {Bazhanov, Dmitry I and Sivkov, Ilia N and Stepanyuk, Valeri S},
  journal   = {Scientific Reports},
  title     = {Engineering of entanglement and spin state transfer via quantum chains of atomic spins at large separations},
  year      = {2018},
  number    = {1},
  pages     = {14118},
  volume    = {8},
  publisher = {Nature Publishing Group UK London},
}

@Article{Hahn1950,
  author    = {Hahn, E. L.},
  journal   = {Phys. Rev.},
  title     = {Spin Echoes},
  year      = {1950},
  month     = {Nov},
  pages     = {580--594},
  volume    = {80},
  doi       = {10.1103/PhysRev.80.580},
  issue     = {4},
  numpages  = {0},
  publisher = {American Physical Society},
  url       = {https://link.aps.org/doi/10.1103/PhysRev.80.580},
}

@Article{Yosida1966,
  author    = {Yosida, Kei},
  journal   = {Phys. Rev.},
  title     = {Bound State Due to the $s\ensuremath{-}d$ Exchange Interaction},
  year      = {1966},
  month     = {Jul},
  pages     = {223--227},
  volume    = {147},
  doi       = {10.1103/PhysRev.147.223},
  issue     = {1},
  numpages  = {0},
  publisher = {American Physical Society},
  url       = {https://link.aps.org/doi/10.1103/PhysRev.147.223},
}

@Article{Doniach1977,
  author   = {S. Doniach},
  journal  = {Physica B+C},
  title    = {The Kondo lattice and weak antiferromagnetism},
  year     = {1977},
  issn     = {0378-4363},
  pages    = {231-234},
  volume   = {91},
  doi      = {https://doi.org/10.1016/0378-4363(77)90190-5},
  url      = {https://www.sciencedirect.com/science/article/pii/0378436377901905},
}

@InCollection{Kroha2017,
  author    = {J. Kroha},
  booktitle = {The Physics of Correlated Insulators, Metals, and Superconductors Modeling and Simulation},
  publisher = {Verlag des Forschungszentrum J{\"u}lich},
  title     = {Interplay of {Kondo} effect and {RKKY} interaction},
  year      = {2017},
  editor    = {E. Pavarini and Erik Koch and R. Scalettar and R. M. Martin},
  volume    = {7},
}

@Article{Allerdt2017,
  author    = {Allerdt, A. and Feiguin, A. E. and Das Sarma, S.},
  journal   = {Phys. Rev. B},
  title     = {Competition between Kondo effect and RKKY physics in graphene magnetism},
  year      = {2017},
  month     = {Mar},
  pages     = {104402},
  volume    = {95},
  doi       = {10.1103/PhysRevB.95.104402},
  issue     = {10},
  numpages  = {8},
  publisher = {American Physical Society},
  url       = {https://link.aps.org/doi/10.1103/PhysRevB.95.104402},
}

@Article{Walls1983,
  author    = {Walls, Daniel F},
  journal   = {nature},
  title     = {Squeezed states of light},
  year      = {1983},
  number    = {5939},
  pages     = {141--146},
  volume    = {306},
  publisher = {Nature Publishing Group UK London},
}

@Article{Wu1987,
  author    = {Wu, Ling-An and Xiao, Min and Kimble, HJ},
  journal   = {Journal of the Optical Society of America B},
  title     = {Squeezed states of light from an optical parametric oscillator},
  year      = {1987},
  number    = {10},
  pages     = {1465--1475},
  volume    = {4},
  publisher = {Optical Society of America},
}

@Article{Pirkkalainen2015,
  author    = {Pirkkalainen, J.-M. and Damsk\"agg, E. and Brandt, M. and Massel, F. and Sillanp\"a\"a, M. A.},
  journal   = {Phys. Rev. Lett.},
  title     = {Squeezing of Quantum Noise of Motion in a Micromechanical Resonator},
  year      = {2015},
  month     = {Dec},
  pages     = {243601},
  volume    = {115},
  doi       = {10.1103/PhysRevLett.115.243601},
  issue     = {24},
  numpages  = {5},
  publisher = {American Physical Society},
  url       = {https://link.aps.org/doi/10.1103/PhysRevLett.115.243601},
}

@Article{Marti2024,
  author    = {Marti, Stefano and von L{\"u}pke, Uwe and Joshi, Om and Yang, Yu and Bild, Marius and Omahen, Andraz and Chu, Yiwen and Fadel, Matteo},
  journal   = {Nature Physics},
  title     = {Quantum squeezing in a nonlinear mechanical oscillator},
  year      = {2024},
  number    = {9},
  pages     = {1448--1453},
  volume    = {20},
  publisher = {Nature Publishing Group UK London},
}

%

\end{document}